\let\iftest=\iffalse

\pdfoutput=1

\documentclass[11pt,a4paper,final]{article}

\usepackage{jheppub}
\usepackage{amssymb,amsmath}
\usepackage{dsfont}
\usepackage{color}
\usepackage{slashed}
\usepackage{wrapfig}

\iftest\else
\PassOptionsToPackage{hypertex}{hyperref}
\PassOptionsToPackage{draft}{graphicx}
\usepackage{showkeys}
\fi

\providecommand{\hypersetup}[1]{}

\hypersetup{plainpages=false}
\hypersetup{pdfpagemode=UseNone}
\hypersetup{bookmarksnumbered=true}
\hypersetup{pdfstartview=FitH}
\hypersetup{colorlinks=false}

\renewcommand{\[}{\begin{equation}}
\renewcommand{\]}{\end{equation}}

\newcommand{\be}{\begin{equation}}
\newcommand{\ee}{\end{equation}}
\newcommand{\bea}{\begin{eqnarray}}
\newcommand{\eea}{\end{eqnarray}}

\begin{document}

\title{Cluster Algebras and the Positive Grassmannian}

\author{Miguel F. Paulos,}
\author{Burkhard U.\ W.\ Schwab}

\emailAdd{miguel\_paulos@brown.edu}
\emailAdd{burkhard\_schwab@brown.edu}

\affiliation{%
Brown University %
(Department of Physics), %
182 Hope St, Providence, RI, 02912%
}


\date{\today}

\abstract{
Plabic graphs are intimately connected to the positroid stratification of the positive Grassmannian. The duals to these graphs are quivers, and it is possible to associate to them cluster algebras. For the top-cell graph of $Gr_{+}(k,n)$, this cluster algebra is the homogeneous coordinate ring of the corresponding positroid variety. We prove that the same statement holds for plabic graphs describing lower dimensional cells. In this way we obtain a map from the positroid strata onto cluster subalgebras of $Gr_{+}(k,n)$. We explore some of the consequences of this map for tree-level scattering amplitudes in $\mathcal N=4$ super Yang-Mills theory.
}

\maketitle

\section{Introduction}

Recently it has been understood that there is a deep connection between the mathematics of the positive Grassmannian and on-shell scattering amplitudes in $\mathcal N=4$ super Yang-Mills theory \cite{ArkaniHamed:2012nw}. This connection translates the beautiful combinatorial and geometric properties of the former into physical principles such as locality and unitarity of the former. The link between the two is possible due to a hidden $GL(k)$ symmetry of scattering amplitudes, which is made manifest by an integral formula over the Grassmannian \cite{ArkaniHamed:2009dn}

The positroid decomposition of the positive Grassmannian is labeled by plabic graphs \cite{2006math......9764P}. Since each cell in the positroid stratification is also a projective algebraic variety via the Pl\"ucker embedding, then to each such graph also corresponds a particular (open) positroid variety. Plabic graphs have been known to be important in physics in a number of areas, and in the connection mentioned above, the graphs are direct representatives of on-shell scattering processes in $\mathcal{N}=4$ SYM theory \cite{ArkaniHamed:2012nw} (for further work, see, e.g., \cite{Beisert:2014qba,Beisert2014,Franco:2014nca,Franco:2013pg,Huang:2013owa,Kanning:2014maa}). Importantly for us, the duals to these graphs are quivers, and it is possible to associate Pl\"ucker coordinates to each node of the quiver in such a way that we obtain a seed for a cluster algebra. In this way we have a map from any given positroid variety onto a cluster algebra. But, there is another natural cluster algebra associated to such a variety, namely its homogeneous coordinate ring. In this paper we show that, as widely expected\footnote{See conjecture 3.4 of \cite{2014arXiv1401.5137M} for a concrete statement, and other interesting properties of cluster algebras associated to positroid varieties.}, these two cluster algebras are the same. 

This statement has already been shown to be true for the top-cell in the positroid decomposition \cite{Scott2003}. The main technical difficulty with lower dimensional cells is as follows. Firstly, there are many equivalent plabic graphs labeling the same variety. These graphs are related to each other by square moves, which at the level of the dual cluster algebra correspond to mutations. Now, going to a boundary in the positroid decomposition is quite simple from the point of view of the plabic graphs -- it corresponds to removing a face by deleting an edge. At the level of the dual quiver, losing a face amounts to deleting a node, which is simple enough -- but if we don't do this carefully, the labels of the other nodes may change in such a way that the resulting cluster algebra is not easily determined. To summarize, whereas with plabic graphs we are typically only interested in the equivalence class of graphs up to various moves, from the point of view of the dual it is important to keep track of precisely which graph one is talking about. To get around this, our main technical tool is a special \emph{freeze-deletion} procedure, which implements going to a boundary of a cell at the level of the cluster algebra in a controlled way. With this procedure we have control over the cluster $a$-coordinates present in the quiver at every step, which allows us to show that the resulting cluster algebra is indeed equivalent to the positroid variety's coordinate ring. This freeze-deletion procedure is essentially the inverse of ``adding a BCFW bridge'' \cite{ArkaniHamed:2012nw}, which guarantees that every cell can be obtained by a sequence of such moves.

This result means that we have a map from the positroid cell decomposition onto the cluster algebra: the top-cell maps onto the full cluster algebra; and lower cells map onto certain preferred subalgebras. For instance, the five codimension-1 boundaries of the top cell of $Gr_{+}(2,5)$ correspond to the five $A_1$ subalgebras of $A_2$ (which is the $Gr_{+}(2,5)$ cluster algebra in the ADE classification); but the six codimension-1 boundaries of the top cell of $Gr_{+}(3,6)$ map onto a  preferred set of six $A_3$ subalgebras of $D_4$, out of a total set of sixteen (twelve $A_3$'s and four $A_1\times A_1\times A_1$'s). This is even more striking for other cluster algebras such as $Gr_{+}(4,8)$, where there is always a finite set of preferred subalgebras out of an infinite set. We investigate this structure in detail for $D_4$ ($Gr_{+}(3,6)$). In particular we consider tree-level scattering amplitudes in $\mathcal{N}=4$ SYM which seem to further single out a further subset of the allowed algebras. Besides their interest for tree amplitudes, we hope that these investigations will act as a stepping stone for an investigation into their appearance in the context of loop amplitudes  \cite{Golden:2013xva,Golden:2014xqa}, where they are responsible for remarkable simplifications in the structure of 2-loop MHV amplitudes.

Here's a quick layout of this note: in the next section we discuss the link between plabic graphs and cluster algebras.  In particular we emphasize that plabic graphs generically do not capture the full algebra but only its cluster algebraic structure. Section \ref{proof} is our main result. We define special deletion and freeze/deletion procedures which allow us to identify the cluster algebras for any cell in the positroid stratification. The following section discusses a few concrete applications and examples, essentially in the context of scattering amplitudes in $\mathcal{N}=4$ super Yang-Mills theory. Finally in the appendix we review the positroid decomposition of the positive Grassmannian, its association with plabic graphs, as well as basic notions of cluster algebras.

\section{Positroid Varieties, Plabic Graphs and Cluster Algebras}

\subsection{Cluster algebras from dual graphs}

\begin{wrapfigure}{r}{0.5\textwidth}
  \centering
  \includegraphics[width=0.48\textwidth]{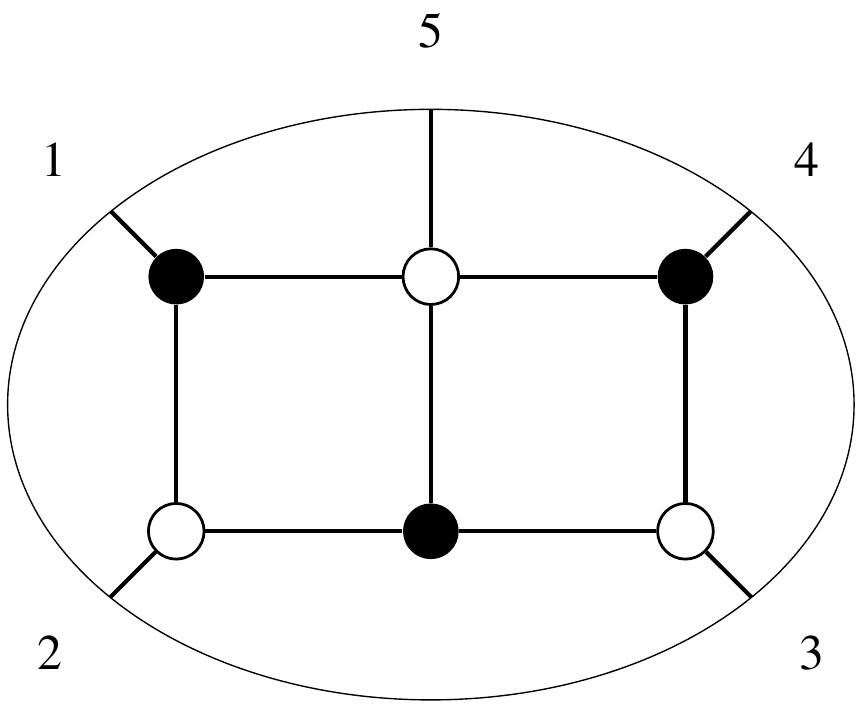}
  \caption{The top-cell graph of $Gr_{+}(2,5)$}
  \label{fig:gr25}
\end{wrapfigure}
Diagram \ref{fig:gr25} is a simple example of a plabic graph. From a physics point of view, it describes the scattering of five particles in the MHV helicity configuration. Mathematically, this graph is associated with the top-cell of $Gr_{+}(2,5)$, and is in the equivalence class of graphs with permutation $\{3,4,5,6,7\}$. There are many other equivalent graphs, obtained by performing square moves, merge/unmerge (explained in apdx.~\ref{sec:shell-graphs-cluster}), which lead to the same permutation; so that really what matters at this point is the equivalence class of all such graphs. To every such equivalence class of plabic graphs one can associate a cluster algebra. Let us briefly review how this is done. 

Firstly, a plabic graph is planar, and so it has a dual graph. In turn this dual graph can be made into a quiver: the orientation of the arrows is determined such that black nodes are always to the right of a particular arrow. Further, if an arrow is dual to an external edge we draw it as a dashed arrow. As a matrix, the quiver has $\pm 1$ entries for solid arrows, and $\pm 1/2$ entries for dashed arrows. Faces in the plabic graph become nodes in the quiver. If a node is dual to an external face we say that it is \emph{frozen}, while interior faces correspond to \emph{mutable} nodes. Finally, in order to obtain the cluster $\mathcal a$-coordinates associated to each node in the quiver, we need to consider the directed left-right paths used to determine the permutation of the graph (see \ref{sec:shell-graphs}). Each of these paths divides the graph into two halves. To every face on the left -- with respect to the direction of the arrow on the path -- of one such path, is associated the number of the leg from which we started. Doing this for every one of the $n$ exterior legs results in all faces being labeled with $k$ integers.  

\begin{figure}[htpb]
\centering
 \includegraphics[width=\textwidth]{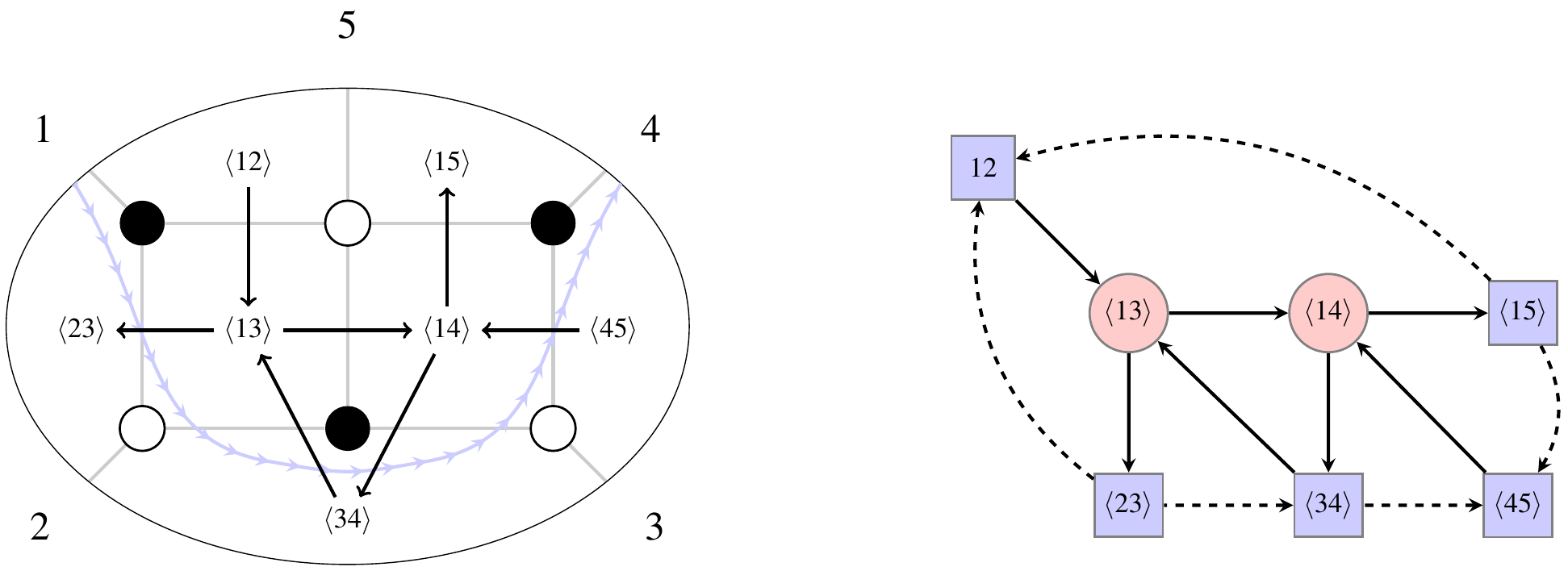}
 \vspace{0.5 cm}
 
  \includegraphics[width=\textwidth]{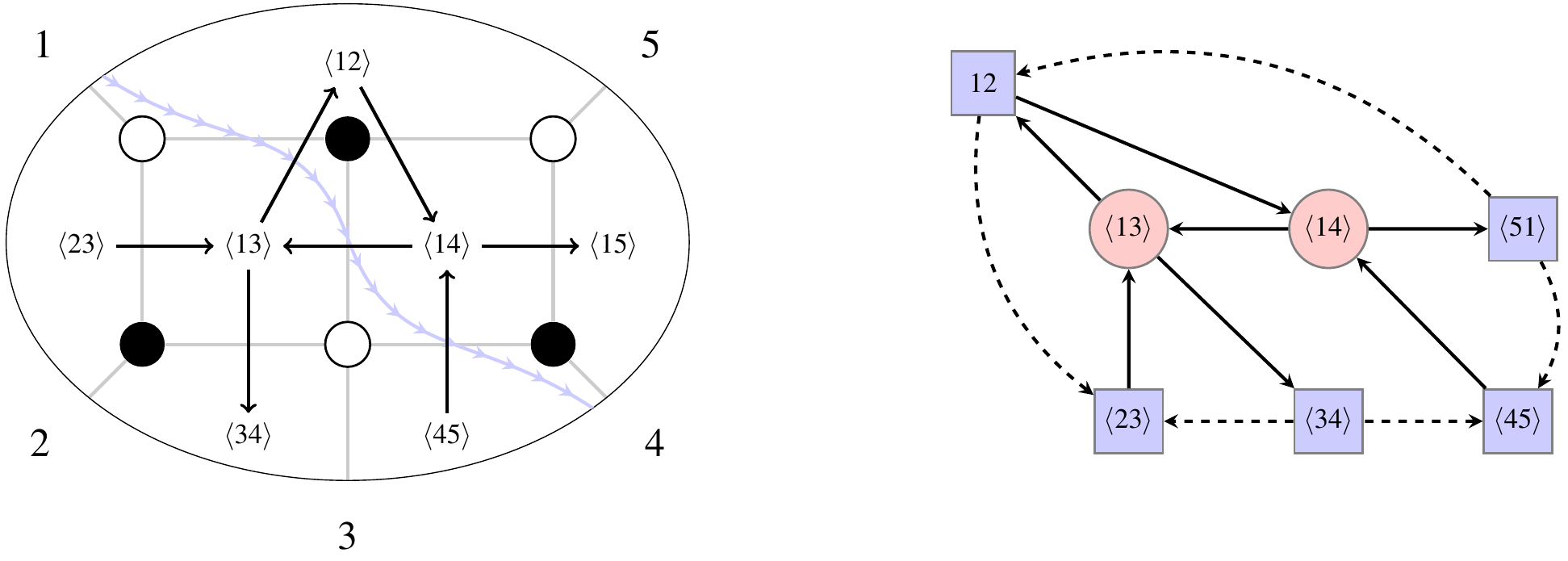}
\caption{Top: a top-cell graph of $Gr(2,5)$ and its dual quiver. Bottom: the same cell after performing a square move and merge-unmerge.}
\label{fig:gr251}
\end{figure}

This procedure is illustrated for the top cell of $Gr_{+}(2,5)$ in figure \ref{fig:gr251}.
%
%
The knowledgeable reader may recognize the quiver as one of those associated to the $Gr(2,5)$ cluster algebra, or $A_2$ in the ADE classification. This not an accident \cite{Scott2003}: for plabic graphs labeling the top-cell in the positroid decomposition of $Gr(k,n)$, the cluster algebra so obtained is precisely the coordinate ring of the top-cell positroid variety, that is, the $Gr(k,n)$ cluster algebra, once we identify the labels of each node as Pl\"ucker coordinates. Under this correspondence each graph corresponds to a seed, and square moves relating different graphs generate mutations in the dual quiver. However, it is not true that all mutations correspond to square moves, as we shall see in next section. Before we do so, it is important to emphasize that the procedure above always yields some cluster algebra given a plabic graph -- but it is not guaranteed {\em a priori} that this cluster algebra should have anything to do with the Grassmannian, and in particular with the coordinate ring of the corresponding positroid variety. That this is actually the case for reduced plabic graphs is precisely what we will show in section \ref{proof}, and the main goal of this note.

\subsection{The accessible algebra}
\label{sec:accessible}

By construction, every quiver dual to a plabic graph will have Pl\"ucker coordinates located on its nodes. However, it is well known that for Grassmannian cluster algebras, the full set of $a$-coordinates generically contains more complicated variables \cite{Scott2003}. How can this be? The solution to the apparent puzzle is that plabic graphs and their duals do not exhaust the set of all possible seeds in the cluster algebra. This is because some seeds can only be obtained by performing mutations on nodes with valency higher than four, and these cannot be seen with plabic graphs -- indeed, such graphs contain hexagonal or higher $2m$-gonal faces, but there are no associated moves. While the mutation rules of the associated cluster algebras allow us to \emph{mutate} on these faces, the result is a non-planar quiver - and accordingly, a dual graph cannot be uniquely associated with it. 

In the following, whenever we talk about the cluster algebra generated by a graph, we mean the algebra generated by \emph{all} mutations, including non-square moves. This will in principle include seeds with non-planar quivers. So, e.g., when we say the top cell graph of $Gr(3,6)$ is associated to $D_{4}$, what we mean is that any representative graph will have for dual a seed belonging to the $D_4$ cluster algebra. By successive mutations one obtains all the seeds in the cluster algebra, only some of which will have plabic graph representatives. Furthermore, all those graphs will be in the same equivalence class, since they will all be related to each other by mutations, and therefore be labeled by the same permutation. At the same time, we denote by \emph{accessible algebra} the part of the cluster algebra that can be reached by only performing square moves in the dual graph -- that is, by mutating only nodes of valency four. This is the part of the algebra that only involves Pl\"ucker coordinates. A particularly important case is the $A_n$ series, which is associated to $Gr(2,n+3)$ -- in this case all seeds contain nothing but Pl\"ucker coordinates, and accordingly every seed has a plabic graph representative.

So far, all these remarks refer to top cell graphs, where we know precisely what exactly is the cluster algebra dual to a plabic graph. In the next section we shall show that generically, plabic graphs corresponding to lower dimensional cells will be associated to certain subalgebras of the cluster algebra corresponding to $Gr_{+}(k,n)$. For each of these, we can only capture those seeds where all cluster $a$-coordinates are Pl\"uckers. But, in the dual cluster algebra we are free to mutate at will, and hence we will still think of a full cluster (sub)algebra being associated to each such graph, as well as a more restricted  accessible part of it.

\section{Freezing and Deletion, and the Positroid Decomposition}
\label{proof}
\subsection{General idea}
We have seen how top-cell graphs, via their duals, have an associated cluster algebra, which is the algebra describing the coordinate ring of $Gr_{+}(k,n)$ - the set of homogeneous polynomials defined on the Grassmannian when we think of it as a projective algebraic variety through the Pl\"ucker embedding. We now want to consider plabic graphs corresponding to lower dimensional cells in the positroid stratification. These graphs may be obtained starting from the top cell by successively deleting certain allowed edges. Combinatorially, we are performing adjacent transpositions on the permutation associated to the on-shell graph. Algebraically, we are considering smaller and smaller positroid varieties, obtained by sending to zero some Pl\"ucker coordinates, or equivalently, by imposing linear relations among columns of the matrix associated to a particular point in the Grassmannian.

\begin{figure}[htpb]
\centering
\includegraphics[width=\textwidth]{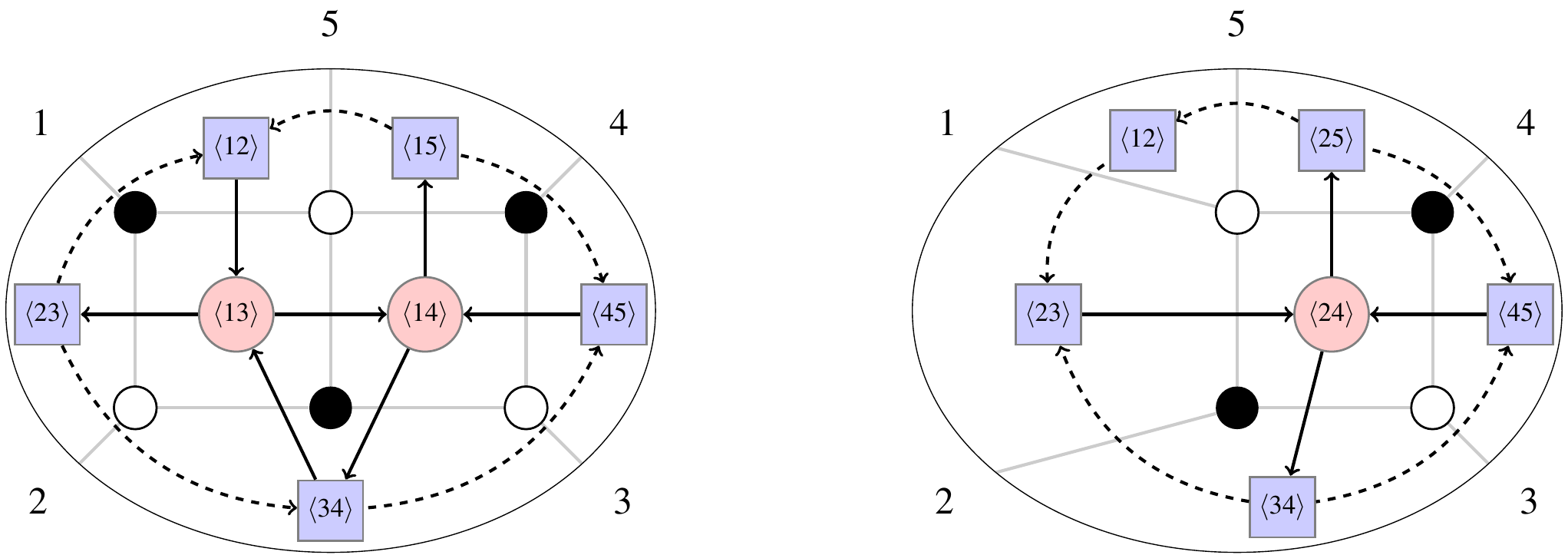}
\caption{Left: a top-cell graph of $Gr(2,5)$ and its dual quiver. Right: a first boundary obtained eliminating an edge.}
\label{fig:gr25bound1}
\end{figure}

Consider figure \ref{fig:gr25bound1}. 
On the left-hand side we have a plabic graph, and its dual quiver, describing the top cell of $Gr(2,5)$. We know that this quiver describes a seed of the $A_2$ cluster algebra, which is the algebra associated to $Gr(2,5)$. On the right-hand side we have a particular codimension-1 boundary of the top cell, obtained by setting the minor $\langle 15 \rangle$ to zero. We can obtain such a graph for instance by cutting the leg joining the two leftmost vertices in the top-cell graph.
Notice how in the right-hand side graph, some of the $a$-coordinates look quite different from those on the left. This is a problem: if we start off with some graph and cut a leg to go to a boundary, it is hard to see what relation if any the resulting cluster algebra should have to the original one. This is in a nutshell what prevents us from identifying the resulting cluster algebra.

\begin{figure}[htpb]
\centering
\includegraphics[width=\textwidth]{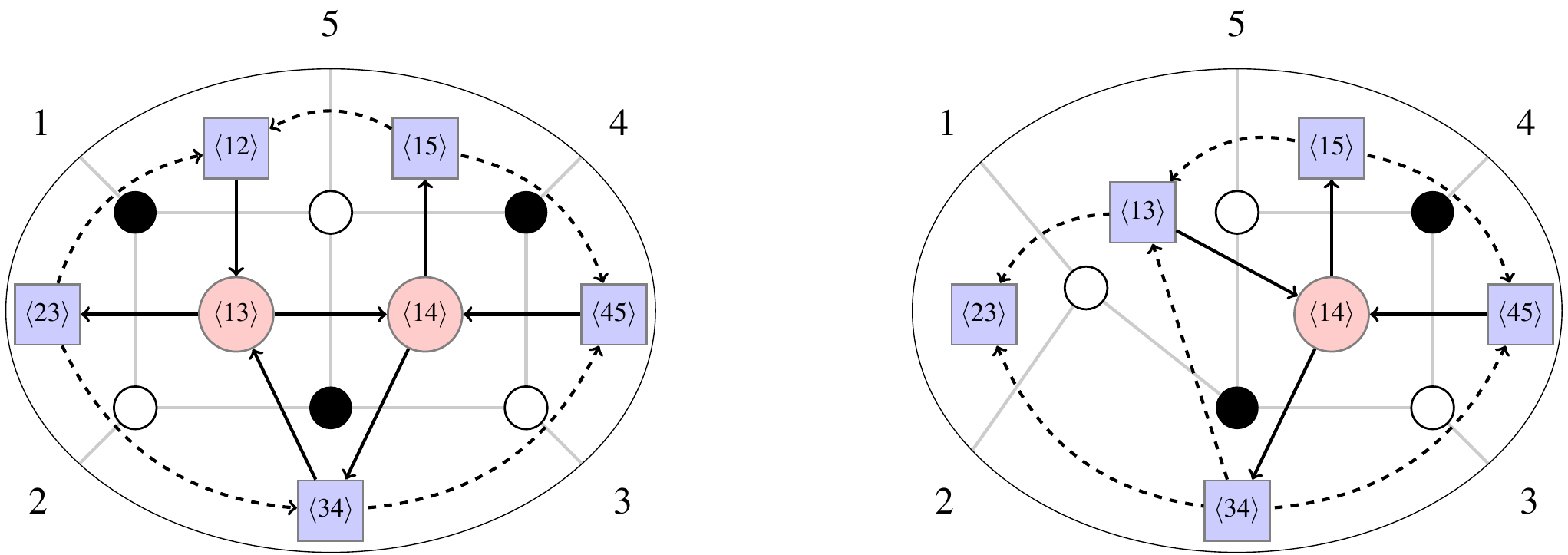}
\caption{Left: a top-cell graph for $Gr(2,5)$, and its dual quiver. Right: a first boundary obtained by setting $\langle 12\rangle\to 0$.}
\label{fig:gr25bound2}
\end{figure}

Consider now figure  \ref{fig:gr25bound2}. 
On the left we have the same top-cell graph, but on the right we have considered a different codimension-1 cell -- that corresponding to setting $\langle 12\rangle\to 0$. As we can see, now the lower dimensional graph has $a$-coordinates which correspond to a subset of the original ones. Indeed we see that two distinct things have happened: we have lost one of the frozen coordinates, $\langle 12 \rangle$, and the coordinate $\langle 13 \rangle$, which was mutable, has taken its place, becoming frozen.   This combined event is what we call a {\em freeze-deletion}. Now, it should be clear that this boundary shouldn't be preferred in any way with respect to the previous one -- how then can it be that we obtain such different results in both cases? The basic reason is that in each case we cut edges with opposite orientations. In the first we have cut an edge with a black node on its left and a white node on its right, and in the second the opposite. If we are to get ``good'' boundaries, we must stick to the latter. But won't this mean that we will miss some of them? The answer of course is no, because there is not one but several distinct top-cell graphs. For instance, we have seemingly singled out leg 5 in the diagrams above. But, performing square moves and merge/unmerge we can rotate the legs cyclically by any desired amount, and for each of those graphs we can take a boundary in the appropriate way. The lesson here then is that it seems to be possible to keep the dual cluster algebra under control by exploiting the fact that one can do square moves on the original graph before taking the boundary.

In our example, going to a boundary involved freezing a mutable vertex, keeping all remaining coordinates the same. Effectively, this restricts us to the set of seeds which contain this vertex. Such ensembles are the codimension-1 subalgebras of the original cluster algebra, which are ensembles of seeds which share at least one $a$-coordinate. In this case, the original algebra is $A_2$, which has five $A_1$ subalgebras. These five subalgebras are in direct correspondence with the five codimension-1 cells of $Gr_+(2,5)$. Indeed, the resulting graphs have a single internal square face, on which we can perform square moves back and forth - generating the $A_1$ algebra. Generically, taking successive boundaries shall involve freezing more and more coordinates -- and accordingly such diagrams will correspond to higher and higher codimension subalgebras of the original cluster algebra.

\subsection{Cluster equivalence}
\label{sec:cluster-equiv}

There is a crucial point we need to make before turning to the general proof: we will take the viewpoint that the dual of a plabic graph does not capture the full algebra $\mathcal A$, but only its cluster algebra structure. This is because it is perfectly possible for two distinct algebras to have the same clusters and mutation rules. Consider for instance a cluster algebra $\mathcal A$ with mutable/frozen cluster variables $\{m_i\}/\{f_j\}$, and define $\mathcal A'\equiv \mathcal A[f_k^{-1}]$, for some $k$. These are clearly two distinct algebras, and yet they share the same clusters \cite{2014arXiv1401.5137M}. This is because from the definition of clusters we must have, for any given cluster $(a_1,\ldots,a_p)$,
\bea
\mathcal A\subset K[a_1^{\pm 1},\ldots,a_p^{\pm 1}].
\eea
Since frozen variables are present in every cluster, then clearly the above is also true replacing $\mathcal A$ by $\mathcal A'$. So $\mathcal A$ and $\mathcal A'$ are both cluster algebras, sharing the same clusters, although they differ as algebras. The same argument also tells us that if we consider $\mathcal A[m_k^{-1}]$, then this will also be a cluster algebra, but one whose clusters are all those clusters in $\mathcal A$ which contain $m_k$. So effectively, from the point of view of $\mathcal A'$, $m_k$ now plays the role of a frozen variable. In this way, this construction can be thought of as a concrete, algebraic way of implementing the freezing procedure described in the previous section. Notice also how in terms of the generalized associahedron of $\mathcal A$, it amounts to considering its codimension-1 faces. In a slight abuse of language we shall think of these as subalgebras of the cluster algebra $\mathcal A$.

\subsection{The proof}

Let us consider a given plabic graph, associated to both a positroid variety $\mathcal V$ and a cluster algebra $\mathcal A$ via its dual. As explained previously, we would like to show that $\mathcal A$ describes the coordinate ring of $\mathcal V$. To do this we can proceed by induction. We start with the top-cell graph for $Gr_{+}(k,n)$, for which this statement is indeed correct \cite{Scott2003}. Now let us consider a plabic graph corresponding to some particular cell in the positroid decomposition, and assume that the statement is true there. Then we shall show that there is a special subset of the boundaries of this cell for which this statement is also true -- boundaries obtained by special freeze and/or deletion procedures defined below. Finally, we shall show that {\em any} cell may be reached from the top-cell in this way.

Given a plabic graph, there are generically two different kinds of boundaries that can be taken. Every time a boundary is taken, we lose an external or internal face of the graph. We consider these two cases in turn.

\subsubsection{Deletion}

\begin{figure}[htpb]
\centering
\includegraphics[width=\textwidth]{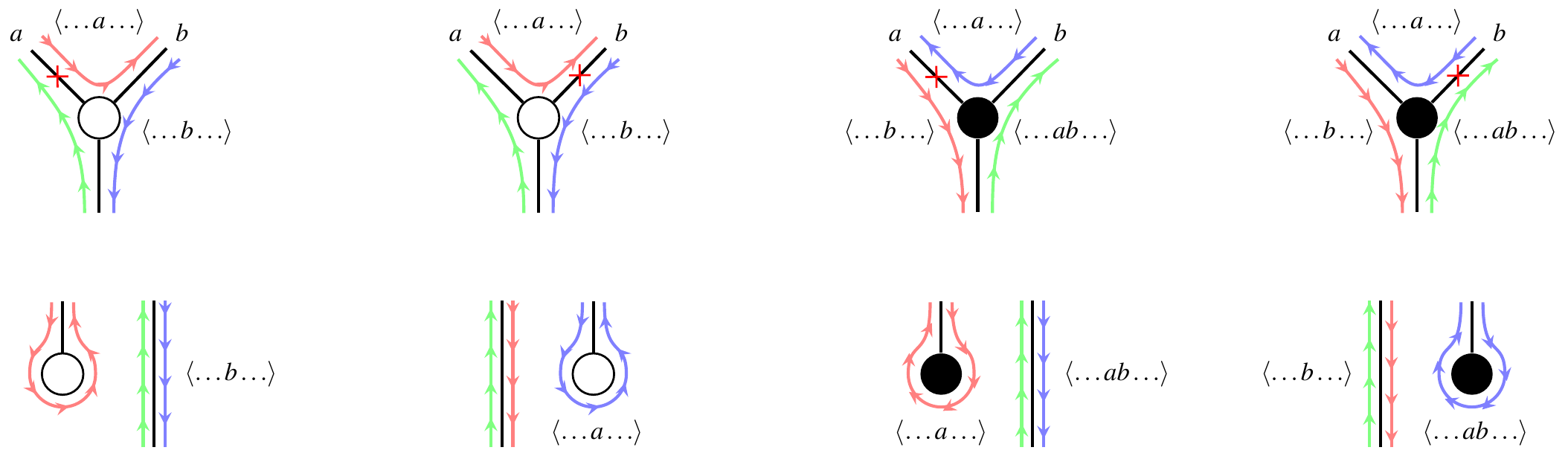}
\caption{Deletions and their effects on the labeling of the nodes.}
\label{fig:del}
\end{figure}

The first case is where we lose an external face, and it is illustrated in figure \ref{fig:del}.
Such boundaries can only be taken if there are two external legs $a,b$ attached to the same vertex, which is something that never occurs for the top-cell graph. We see that there are four distinct cases to consider, depending on whether the legs are attached to a white or black vertex, and which of the two legs we detach. The effect of this on the dual quiver can be understood from the left-right paths, since these are the ones responsible for labeling the nodes of the quiver. Firstly, in any case we lose a frozen node - i.e. the dual to an external face. Secondly, there are two things that may happen to the other nodes: either their labels remain completely unchanged, or labels $a,b$ are swapped. In the case where both legs are attached to a white vertex, everything remains unchanged if we detach the left-most leg; for a black vertex the right-most. Therefore, if we consider only these particular boundaries, the cluster algebra of the dual graph is easily determined:
\bea
\mathcal A \quad \stackrel{\mbox{\small delete}}{\longrightarrow} \quad \mathcal A/I(d)
\eea
where $I(d)$ is the ideal generated by coordinate $d$, which has been sent to zero. Dividing by this ideal effectively sets the variable $d$ to zero. This algebra is by definition the coordinate ring of the positroid variety associated to this boundary, so the result holds in this case. Had we picked the other case, where we detach the opposite leg, this would no longer be true. These choices are not arbitrary, since they lead to distinct boundary cells -- but, as will be explained below, this is not a problem since we'll be able to access any cell by taking only these preferred boundaries. This is only possible because any particular cell with codimension larger than one is always the boundary of two distinct higher dimensional cells \cite{ArkaniHamed:2012nw,MR2350779}, and our particular choice is always correct for one of them.

\subsubsection{Freeze/Deletion}
\begin{figure}[htpb]
\centering
\includegraphics[width=13 cm]{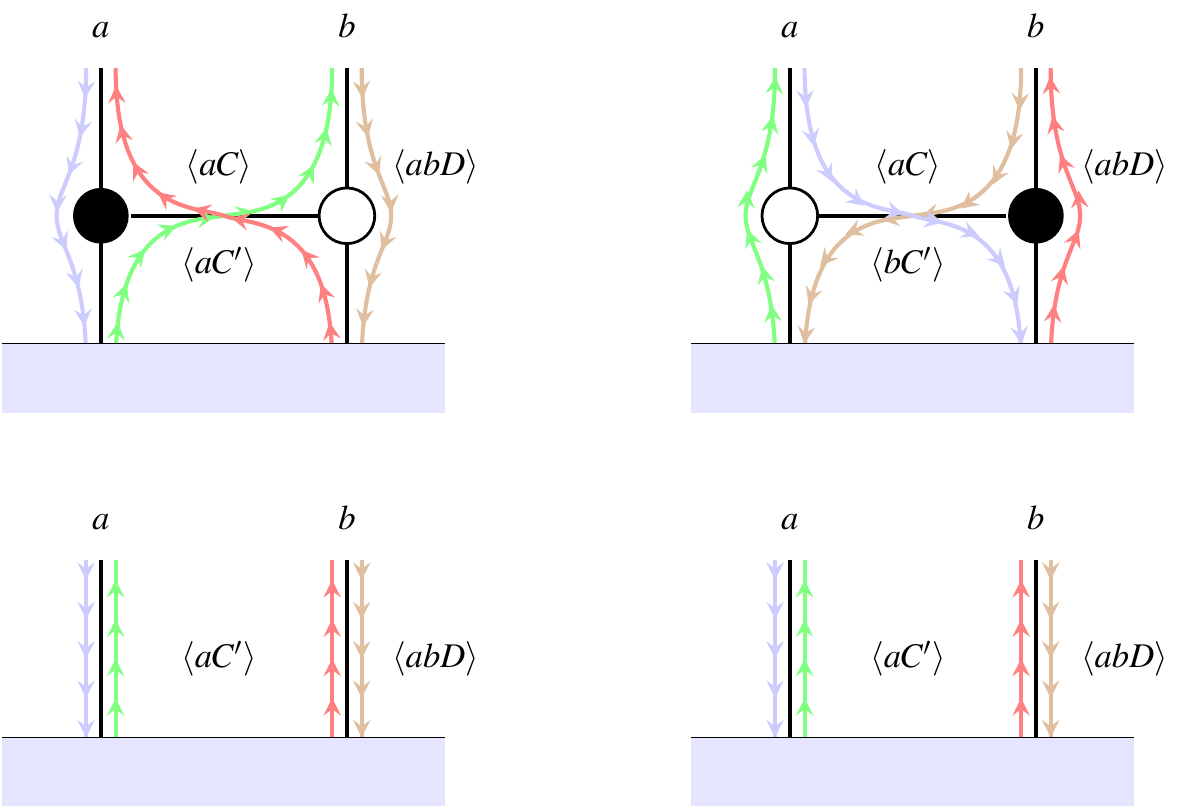}
\caption{Deletion of the edge on the left will have no influence on the entries of the dual quiver. On the right it will induce a swap $a\leftrightarrow b$.}
\label{fig:bridge}
\end{figure}

Let us now move on to the situation where we remove an internal face. We start by considering the dual quiver $Q$ to the plabic graph, and define:
\begin{itemize}
\item {\bf Freezable/Deletable pairs}\\  {\em If a quiver $Q$ has a frozen node with a single outgoing solid arrow, which furthermore connects to an unfrozen node} m {\em then} (m,f) {\em is a freezable/deletable pair}.
\end{itemize}
%
%
The arrow between a freezable/deletable $(m,f)$ pair is dual to an edge separating an external face from an internal one in the plabic graph. Removing this edge removes an internal face, and it corresponds to going to a particular boundary in the cell decomposition. Accordingly, we get a new plabic graph associated to some smaller variety $\mathcal V'$. This boundary is precisely the one which is obtained by sending the Pl\"ucker coordinate $f$ to zero. As shown in figure \ref{fig:bridge}, deleting this edge has the effect of turning $m$ into a frozen node (one that is dual to an external face) and deleting node $f$ from the quiver, and no other change. 
Since this freeze/deletion procedure doesn't affect the other variables in the quiver, we can safely identify a possible cluster algebra that is associated to it - that is, one that has the same clusters: 
\bea
\mathcal A \quad \stackrel{\mbox{\small freeze}}{\longrightarrow} \quad 
\mathcal A[m^{-1}] \quad
\stackrel{\mbox{\small delete}}{\longrightarrow}\quad   \frac{\mathcal A[m^{-1}]}{I(f)}
\eea
The first step follows from our remarks in section \ref{sec:cluster-equiv}. As for the second one, it is the meaning of implementing deletion, just like in the previous section. Now we simply notice
\bea
\frac{\mathcal A[m^{-1}]}{I(f)}=\left(\frac{\mathcal A}{I(f)}\right)[m^{-1}]\simeq \frac{\mathcal A}{I(f)}
\eea
where $\simeq$ means cluster equivalence. $\mathcal A/I(f)$ is exactly the coordinate ring of the variety $\mathcal V'$, so this is precisely what we want. To show the second step we need to prove that every cluster in $\mathcal A/I(f)$ contains $m$ - that is, $m$ is a frozen variable in $\mathcal A/I(f)$. That this is the case follows from the fact that node $m$ was connected to $f$, and $f$ has been sent to zero. So, the possible mutation rule of $m$ becomes
\bea
m' m=\prod_{j=1}^p a_j^{\mbox{\tiny min}(0,-q_{ij})}.
\eea
Due to the Laurent phenomenon \cite{MR1887642}, $m'$ cannot have a pole when $m$ is sent to zero. This implies that the right-hand side is proportional to $m$, and by homogeneity the only possibility is that it is actually trivially equal to the left-hand side. So there is no mutation rule possible, $m$ is indeed a frozen variable in $\mathcal A/I(f)$, and cluster equivalence holds -- and we have precisely the desired result.
\subsubsection{Completeness}

\begin{figure}[htpb]
\centering
\includegraphics[width=\textwidth]{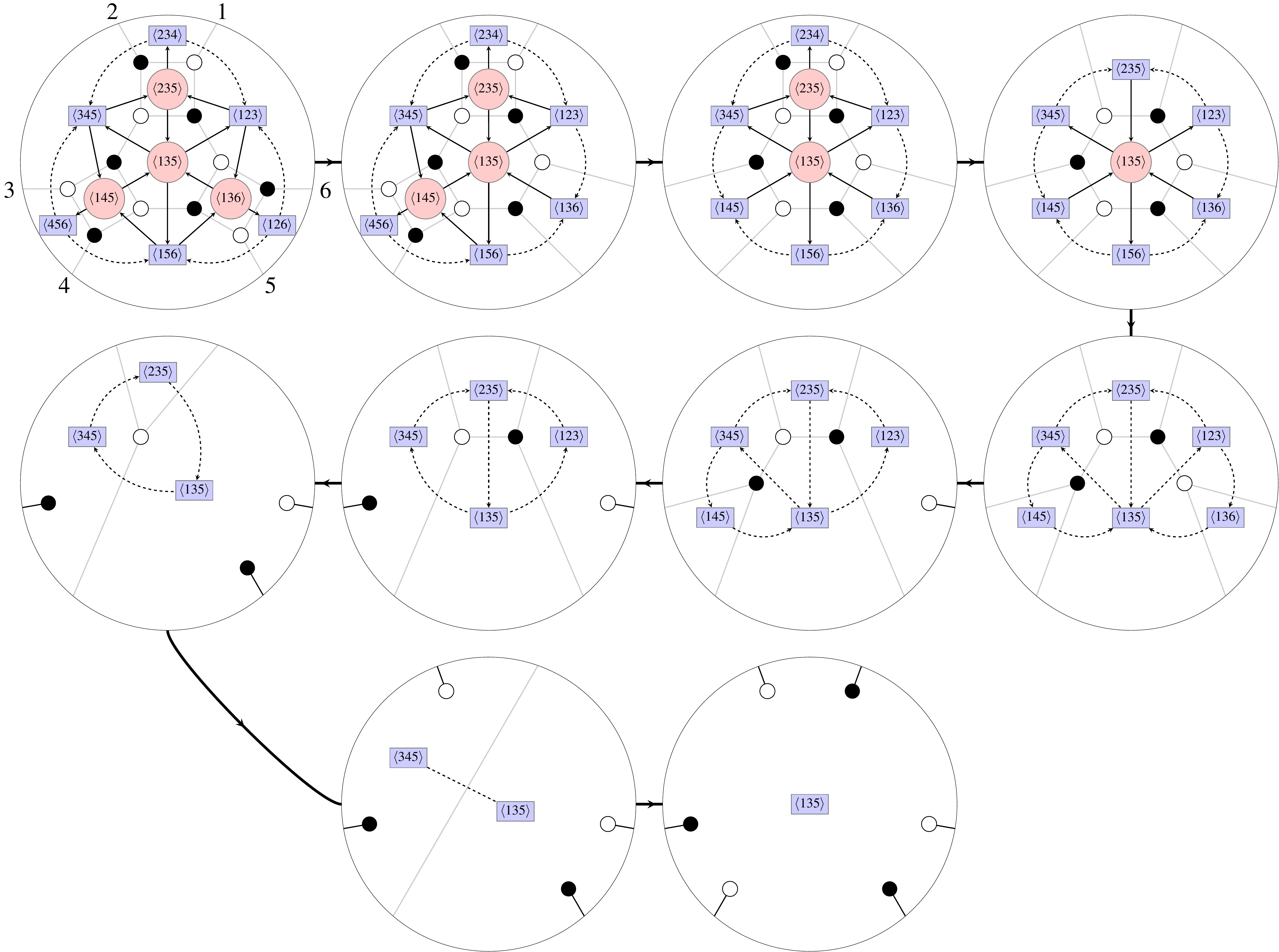}
\caption{A possible path from top-cell to a dimension-1 cell showing our special deletion and freeze/deletion procedures in action. The first four boundaries are obtained via freeze/deletion, the remainer by simple deletion. At every step we lose a node, but the remaining quiver is unchanged. By considering all such paths we can determine the cluster algebra associated to each cell. In this example we start with the $D_4$ cluster algebra (the top cell), move on to an $A_3$ subalgebra, followed by an $A_2$ contained in that $A_3$ and finally an $A_1$ inside $A_2$. The remaining cells are uninteresting from a cluster algebra perspective.}
\label{fig:path}
\end{figure}

To complete our proof, we must show that every cell in the positroid decomposition may be accessed by freeze/deletion or the special deletions considered above. But this can be done by simply noticing that so defined, these procedures are nothing but the reverse of adding a ``BCFW bridge'' \cite{ArkaniHamed:2012nw} to the graph. Every time we remove such a bridge, we perform a transposition on the permutation labeling the graph. Since any permutation is a sequence of transpositions, it follows that we can always take a sequence of freeze/deletions or deletions to reach any particular cell. We exemplify this in figure \ref{fig:path}, where a full path from the top-cell of $Gr(3,6)$ is taken down to the one-dimensional cell, with the dual quiver under complete control at every step.

\section{Some simple applications}

It's time to illustrate the results of the previous sections with a few concrete cases. This will also serve as an introductory exploration of possible applications of these results to physics, and in particular to scattering amplitudes in $\mathcal N=4$ SYM. So, a lightning review of the links between the latter and plabic graphs is in order. We refer the reader to the extensive \cite{ArkaniHamed:2012nw} for more details and further references.

Plabic graphs have the interpretation of on-shell diagrams, made up by gluing together on-shell three-point scattering amplitudes. In our conventions a white (black) vertex corresponds to a three-point MHV ($\overline{\rm MHV}$) amplitude. The numbers $k,n$  of the plabic graph then refer to the N$^{k-2}$MHV level and the number $n$ of external legs of the scattering amplitude. All tree-level amplitudes generically correspond to sums of leading singularities, associated to collections of graphs as determined by the BCFW construction. These leading singularities may be computed either by directly multiplying all the on-shell amplitudes together, or equivalently via the Grassmannian residue formula \cite{ArkaniHamed:2009dn}. Since (reduced) plabic graphs are labeled by permutations, this map implies that leading singularities in turn label cells in the positroid decomposition. For example, the six point MHV amplitude is represented by the unique $k=2$ graph labelled by the permutation \[\binom{1\ 2\ 3\ 4\ 5\ 6}{3\ 4\ 5\ 6\ 7\ 8}.\]
More generally, tree amplitudes will be sums of terms corresponding to several cells. Each such term is also called a BCFW channel. In general we have \cite{ArkaniHamed:2012nw} that for the $n$-point (N)$^{k-2}$MHV tree amplitude,
\[\#{\rm BCFW\ terms\ in\ } A_n  = \frac{1}{n-3}\binom{n-3}{k-1}\binom{n-3}{k-1}.\]
In particular for MHV amplitudes the number of graphs is always one, namely the top-cell graph of $Gr(2,n)$.

A top cell graph has $k(n-k)+1$ faces and therefore as many integrals in the Grassmannian formula. A BCFW channel has $2n$ $\delta$-distributions. Since we need to be left with $4$ $\delta$s for momentum conservation after the calculation, there should be $n_{f}=2n-4 + 1$ faces in each channel. This means that the BCFW channels are recovered after taking  $k(n-k)+4-2n$  boundaries. It also means that every BCFW channel has exactly $n-3$ interior faces independent of $k$.

\subsection{MHV amplitudes}
\label{ss:mhv}
The number $n-3$ is also the number of interior faces of the unique $n$-point MHV graph. In the latter case, this graph generates a cluster algebra of type $A_{n-3}$ \cite{Scott2003}. This, together with the results in the previous section, suggests that generically the BCFW terms that make up tree-level amplitudes could be associated to $A_{n-3}$ cluster (sub)algebras. In the MHV case, these are nothing but the $Gr(2,n)$ cluster algebras, and there is accordingly a single BCFW term. For higher $k$ however, $Gr(k,n)$ will have many possible $A_{n-3}$ subalgebras - in fact, generically there will be an infinite number of them. The positroid decomposition itself selects a restricted, finite number of subalgebras. Tree amplitudes correspond to a further refinement of this selection, as perhaps will become clearer in the next section.

\begin{figure}[htpb]
\centering
\includegraphics[width=6 cm]{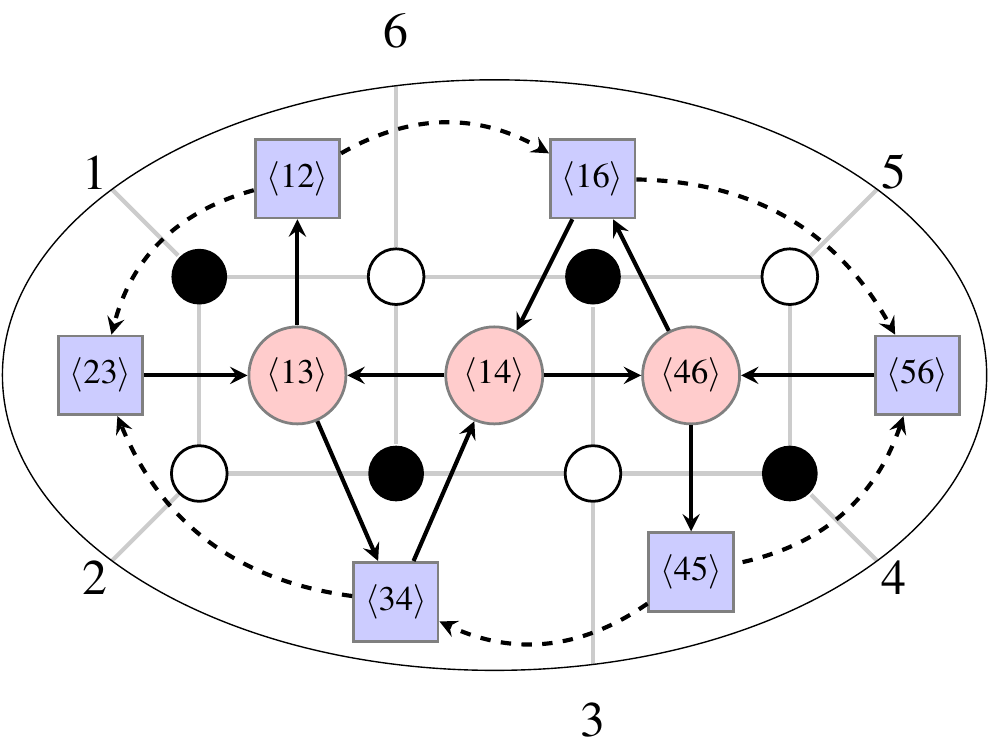}
\caption{A $Gr_{+}(2,6)$ top-cell graph and the corresponding quiver.}
\label{fig:a3topcell}
\end{figure}

\begin{figure}[htpb]
\centering
\includegraphics[width=6 cm]{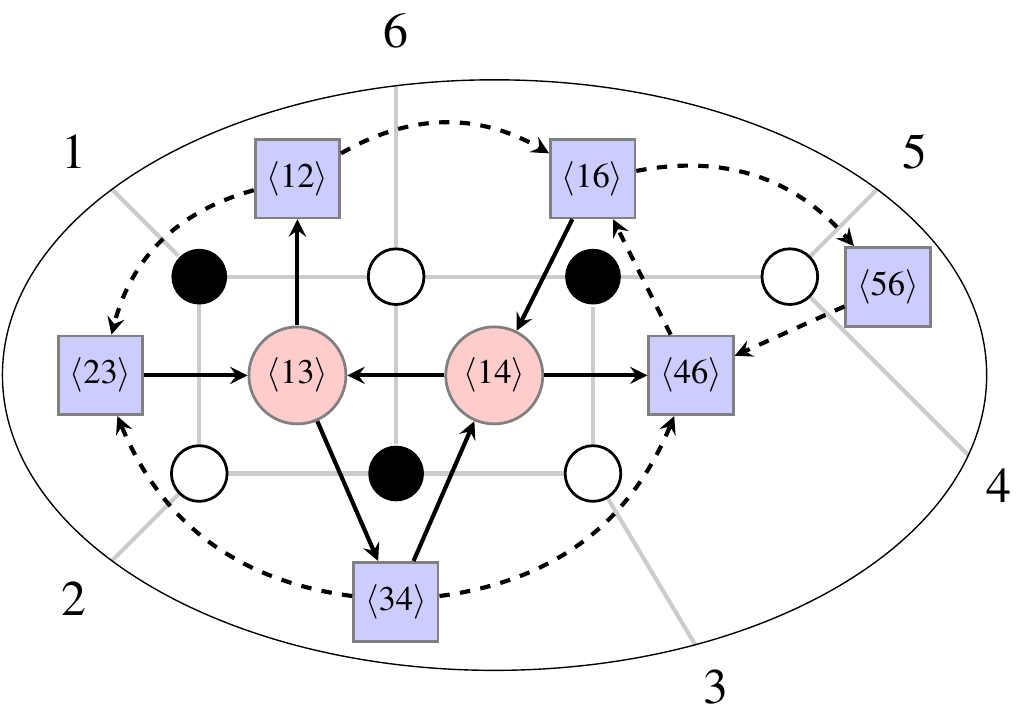}
\caption{A codimension-1 cell of $Gr_{+}(2,6)$, obtained by setting $\langle 45\rangle$ to zero.}
\label{fig:a3bound}
\end{figure}

But first, let us consider the MHV case in more detail. The top cell graph -- along with its dual quiver -- is shown in figure \ref{fig:a3topcell}.
The associated cluster algebra is $A_3$, whose generalized associahedron is shown in figure \ref{fig:a3poly}. In a sense, this object represents the full amplitude, and we see that there is a lot of structure in it -- {\em e.g.} there are six pentagonal faces bounded by three squares. To probe this structure, we can for instance consider taking a collinear limit, aligning particles $i$ and $i+1$. In this limit the Pl\"ucker coordinate $\langle i i+1\rangle$ vanishes, so that we go to a particular boundary of the top cell. We can do this in six particular ways, and in each case we end up with a graph similar to that of figure \ref{fig:a3bound}, where we have sent $\langle 45\rangle$ to zero. 
By our freeze deletion procedure, we know that each of these boundaries can be accessed by freezing a particular $a$-coordinate, or equivalently selecting out a particular codimension-1 subalgebra of $A_3$. $A_3$ has 15 $a$-coordinates, among which $6$ are frozen and 9 unfrozen. Each of the 9 corresponds to a subalgebra, which geometrically is a $2$-face of the generalized associahedron. Three of these are the square faces, which are $A_1\times A_1$ algebras; the remaining six are the six pentagonal faces which span $A_2$ subalgebras. It is these which are captured by the positroid decomposition. Indeed the graph shown in figure \ref{fig:a3bound} has two internal faces, and its dual quiver does generate indeed an $A_2$ algebra.

Proceeding further, we can now take a soft limit, by eliminating one of the legs altogether. That is, having taken legs 4 and 5 to be proportional, we take the proportionality parameter to zero, eliminating leg 5 altogether. This would correspond to our deletion procedure in figure \ref{fig:a3bound}. The result is again an $A_2$ algebra, the same as before, but with the extra frozen coordinate $\langle 56\rangle$ removed. In this way we recover precisely the top-cell graph of $Gr(2,5)$, albeit with legs labeled $1,2,3,4,6$, which is the graph that gives us the five point MHV amplitude for those particles. In this way, we see that quite beautifully the $A_3$ cluster algebra automatically knows something about collinear and soft limits, and the generalized associahedron represents this information in a geometric way. This statement generalizes straightforwardly to the generic $A_n$ cluster algebra.

\subsection{$D_4$ and the 6 pt NMHV amplitude}

\subsubsection{Top-cell}
\begin{figure}[htpb]
\centering
\includegraphics[width=6 cm]{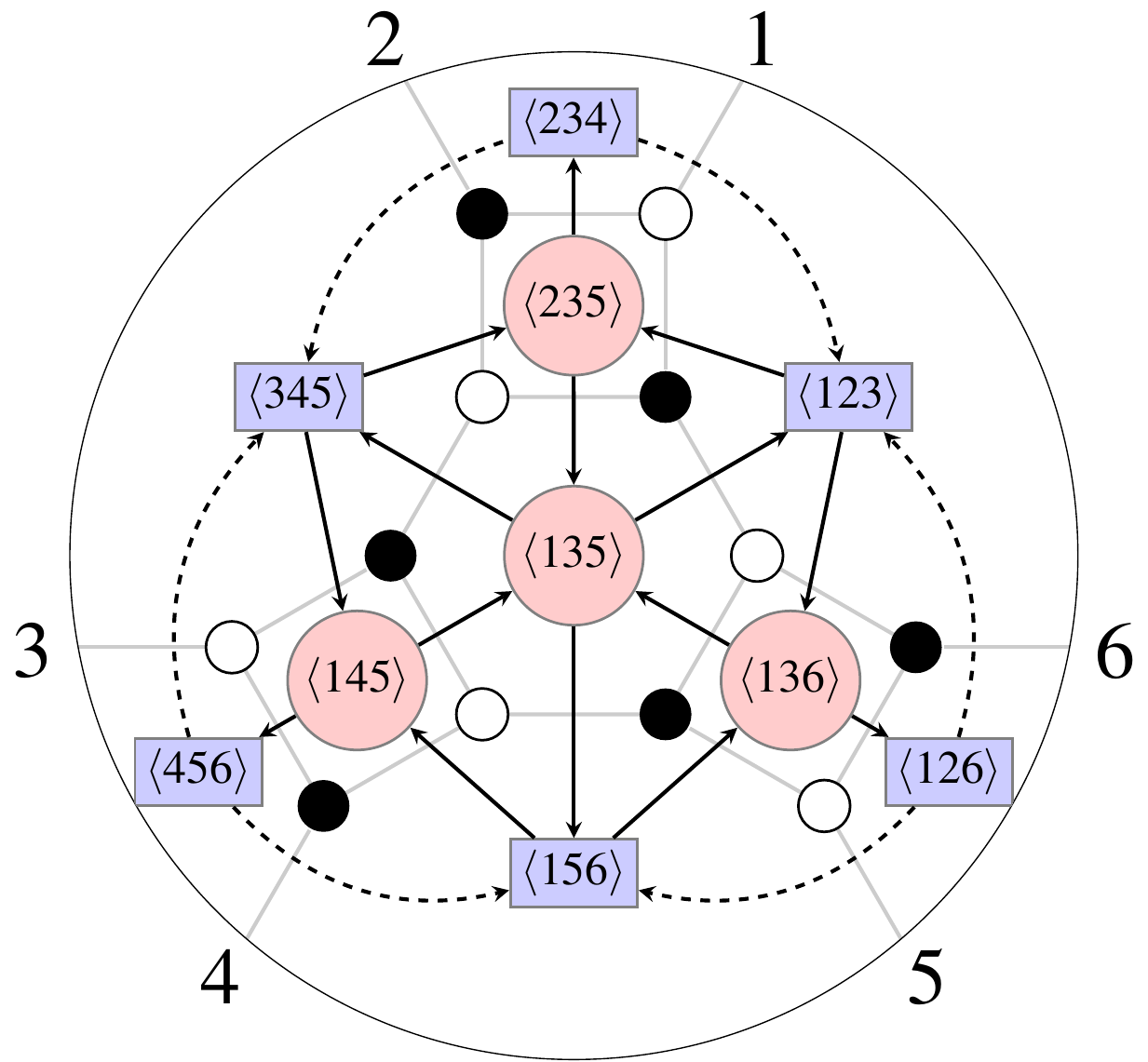}
\caption{A $Gr(3,6)$ top-cell graph and the corresponding quiver.}
\label{fig:gr36topcell}
\end{figure}

We will now examine in detail the particular case of the 6 pt NMHV amplitude. The amplitude can be written as a sum of BCFW terms corresponding to plabic graphs living on the codimension-1 boundary of the $Gr_+(3,6)$ top cell. All such boundaries have three internal faces, and as such we expect that these correspond to $A_3$ subalgebras of the $D_4\equiv Gr(3,6)$ cluster algebra. To construct the $D_4$ cluster algebra itself, we begin with the seed shown in figure \ref{fig:gr36topcell} together with its dual plabic graph.
We can mutate this seed to generate the full cluster algebra. As we have explained in section \ref{sec:accessible}, not all seeds will correspond to plabic graphs, which single out a preferred subset - those seeds which contain only Pl\"ucker coordinates -- and only allow mutations on 4-valent nodes. The best way to visualize this is perhaps to consider the generalized associahedron of $D_4$, and single out those seeds which have dual plabic graphs. 

\begin{figure}[htpb]
\centering
\includegraphics[width=5 cm]{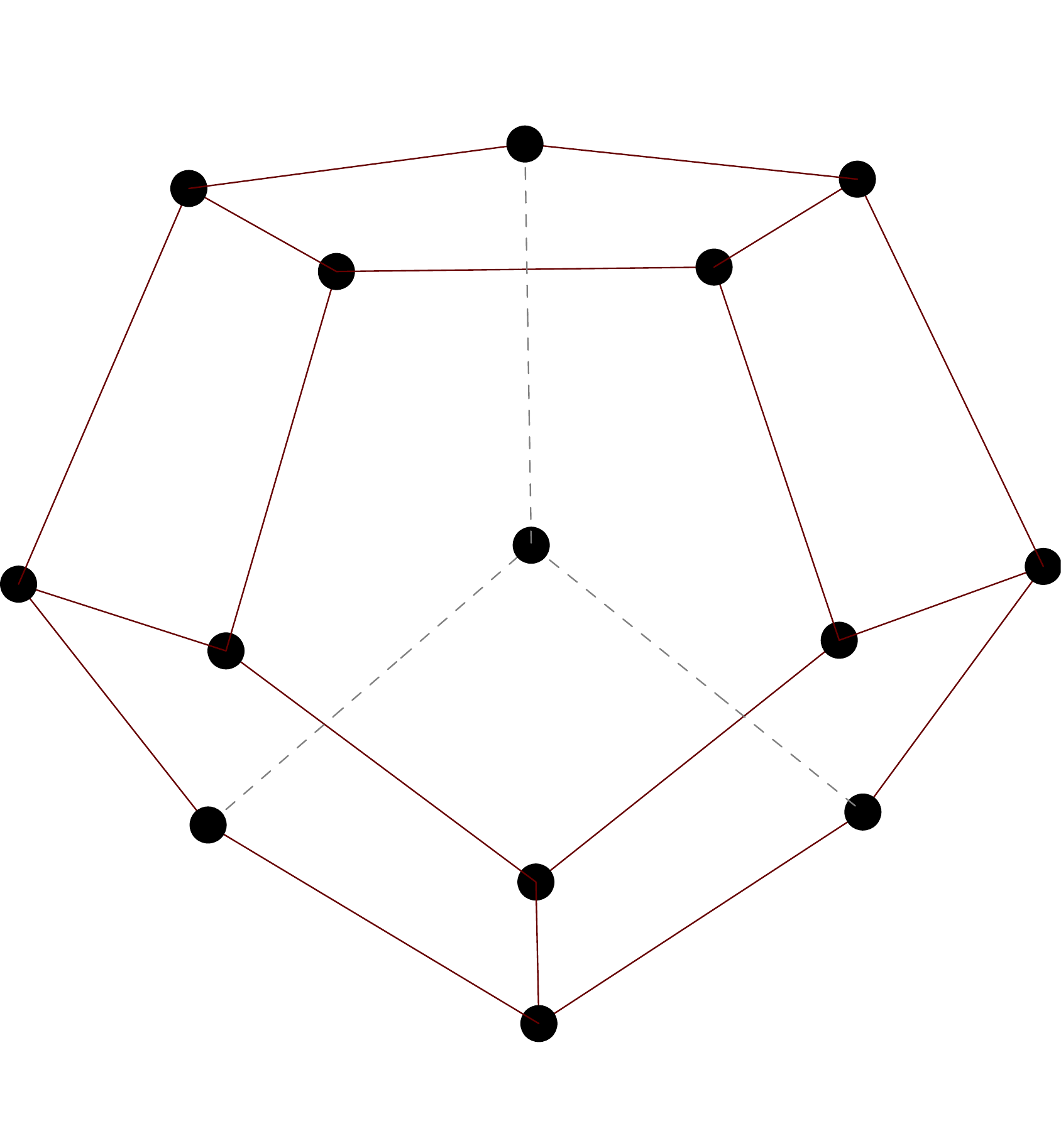}
\caption{The generalized associahedron for the $A_3$ cluster algebra, also known as the Stasheff polytope.}
\label{fig:a3poly}
\end{figure}

\begin{figure}[h]
  \centering
  \begin{tabular}{cc}
   \includegraphics[width=0.4\textwidth]{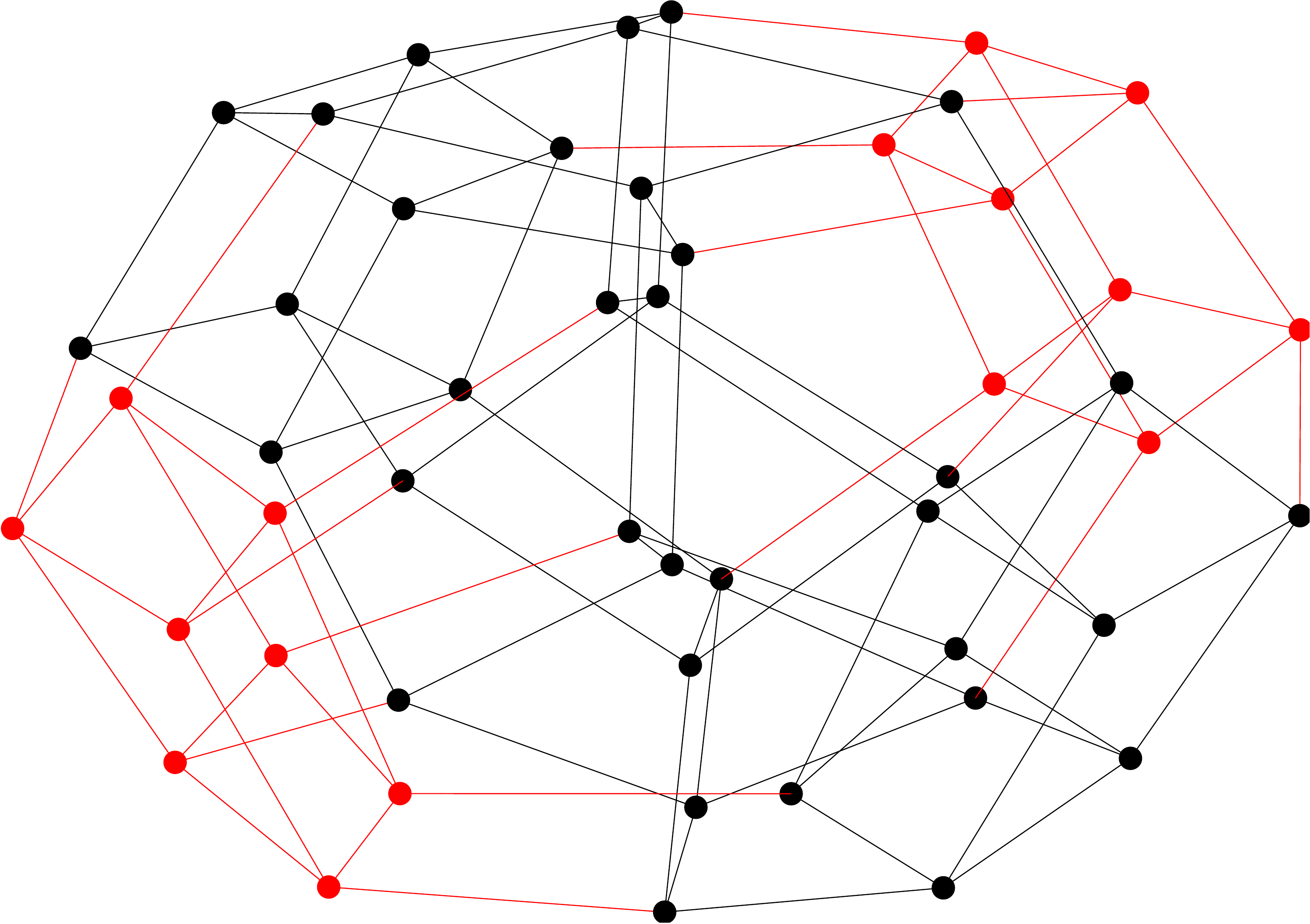} &\hspace{1cm}
  \includegraphics[width=0.4\textwidth]{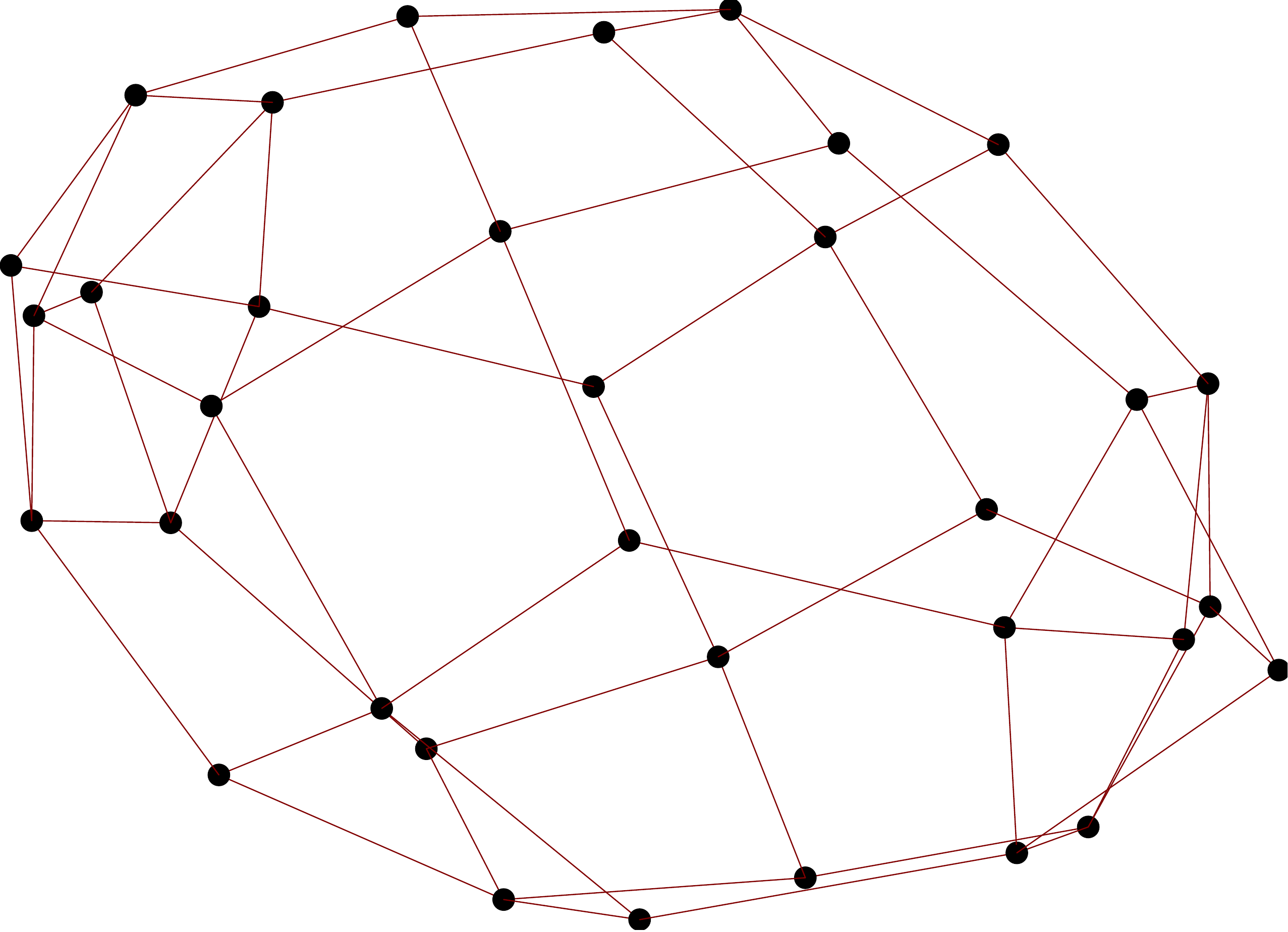} 
    \end{tabular}
  \caption{On the left the $D_4$ generalized associahedron. It is a four dimensional polytope whose three dimensional faces are four cubes and twelve Stasheff polytopes, as in figure \ref{fig:a3poly}. The connections between these faces are better understood via the dual cluster complex in figure \ref{fig:duald4}. The accessible seeds and allowed mutations are shown in black. We display them independently on the right for clarity. They form a curious figure resembling a Faberge' egg.}
  \label{fig:egg}
\end{figure}

\begin{figure}[htpb]
  \centering
  \includegraphics[width=6 cm]{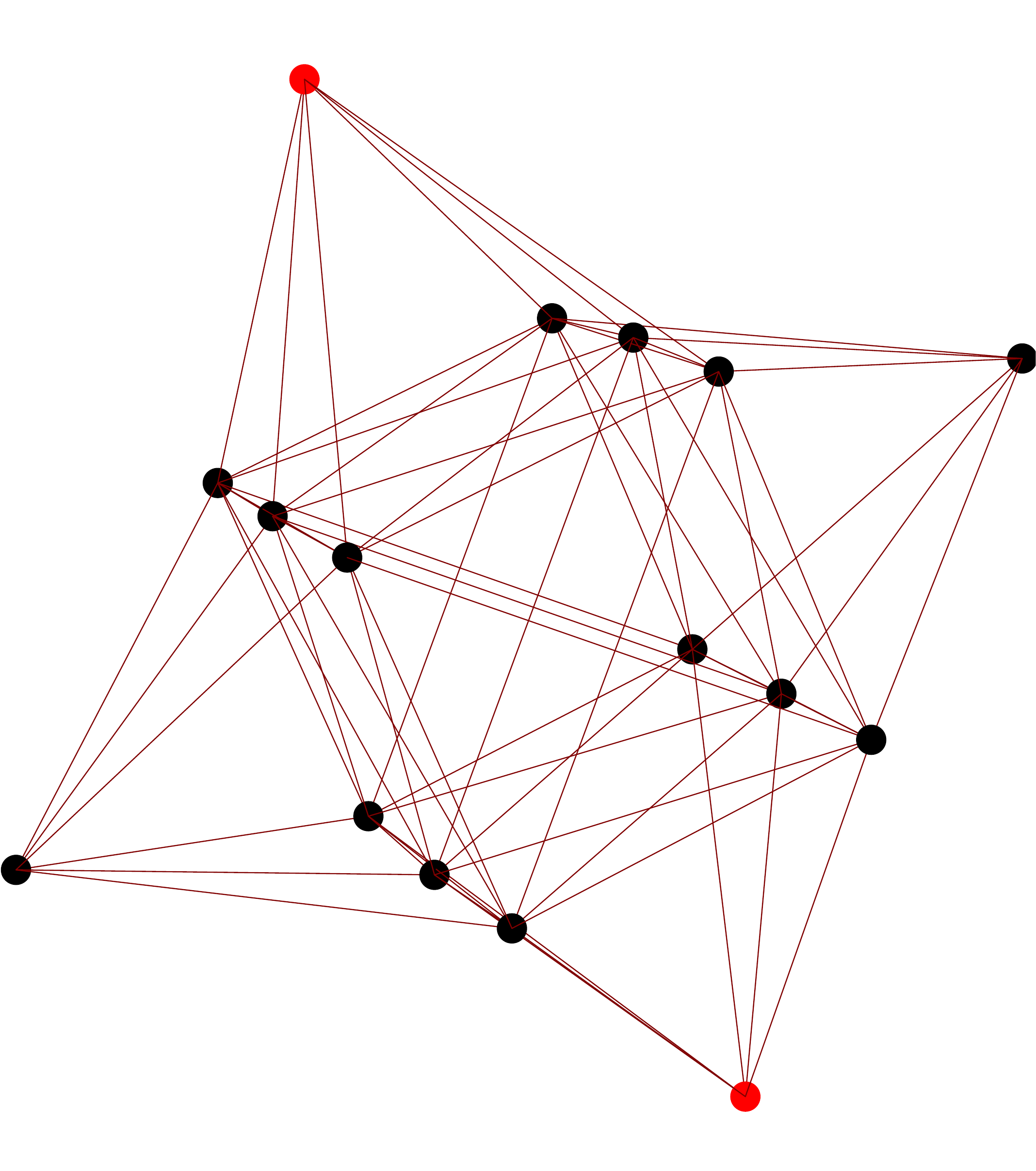}
  \caption{The cluster complex of Gr(3,6). There are 16 nodes, corresponding to the 16 ${\cal A}$-coordinates. Four of these correspond to $A_1\times A_1 \times A_1$ subalgebras. They are the four external nodes. The remaining twelve correspond each to an $A_3$ subalgebra. The two nodes in red represent the two inaccessible A-coordinates, which are cubes in the generalized associahedron.}
  \label{fig:duald4}
\end{figure}

This is shown in figure \ref{fig:egg}. We observe that the accessible polytope (henceforth dubbed the Faberg\'e egg) is clearly not a generalized associahedron by itself since it contains vertices of different valency. 
%
The keen-eyed observer might notice that the missing seeds (red on the lefthandside of figure \ref{fig:egg}) take the shape of two cubes. This is not an accident. The $D_4$ cluster algebra has a total of 16 A-coordinates. To each of these there corresponds a codimension-1 face, or three-dimensional polytope, in the generalized associahedron. These faces correspond to twelve $A_3$ subalgebras, and four $A_1\times A_1 \times A_1$. The cluster complex of $D_4$, which is the dual of the generalized associahedron, is shown in figure \ref{fig:duald4} -- it represents how these faces, which are now represented by vertices, are connected to each other.

The $A_3$ subalgebras each look geometrically as in figure \ref{fig:a3poly}, whereas the $A_1\times A_1\times A_1$ take the shape of cubes. The two missing cubes corresponds precisely to the only two $a$-coordinates which are {\em not} Pl\"ucker coordinates \footnote{They are quadratic in Pl\"ucker coordinates.}, and which therefore are inaccessible from the point of view of plabic graphs.

\subsubsection{Boundaries and the tree amplitude}

\begin{figure}[htpb]
\centering
\includegraphics[width=6 cm]{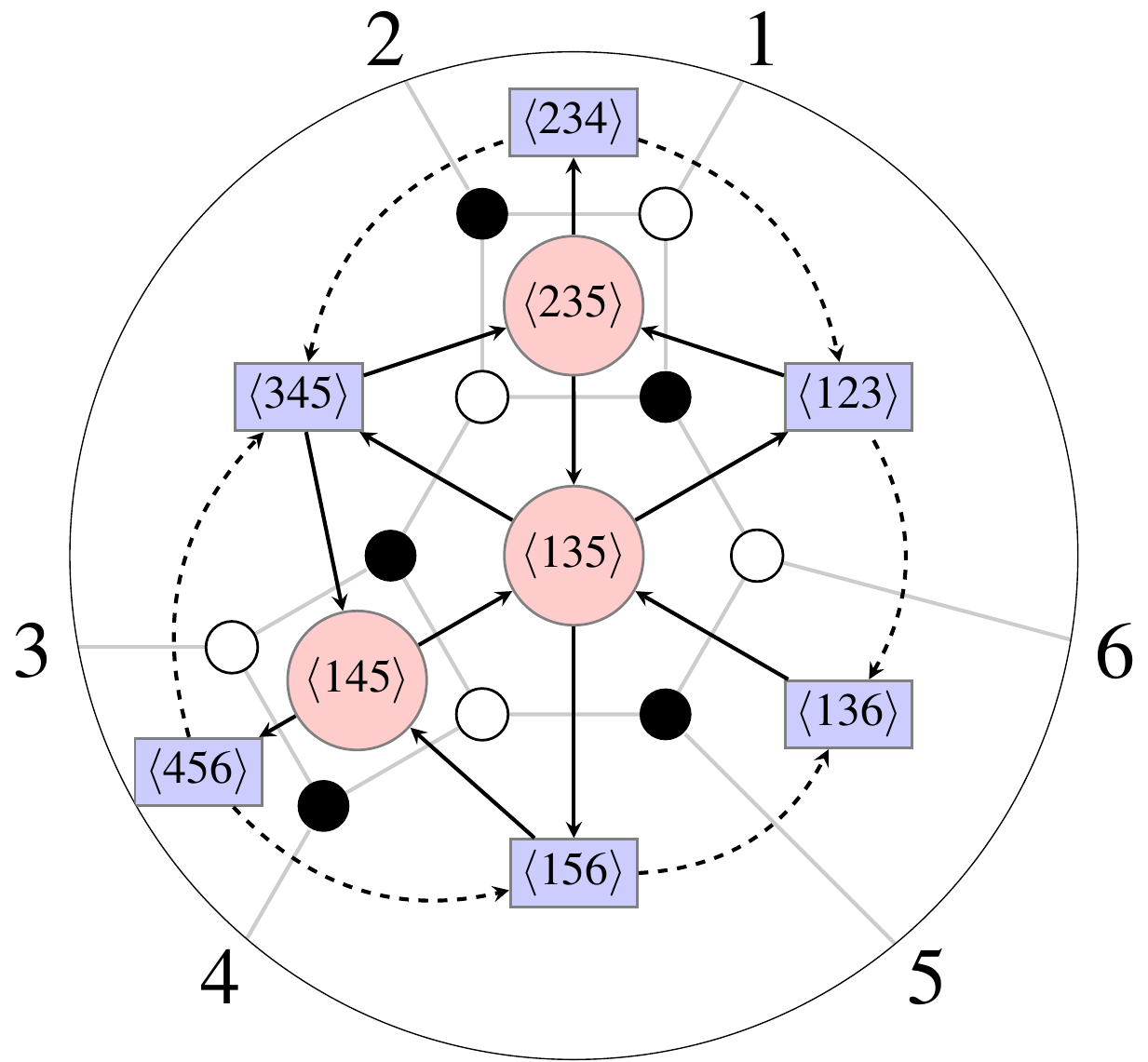}
\caption{A codimension-1 cell of $Gr_{+}(3,6)$, obtained by setting  $\langle 456\rangle$ to zero.}
\label{fig:gr36bound}
\end{figure}

\begin{figure}[htpb]
  \centering
 \begin{tabular}{cc}
  \includegraphics[width=6.5cm]{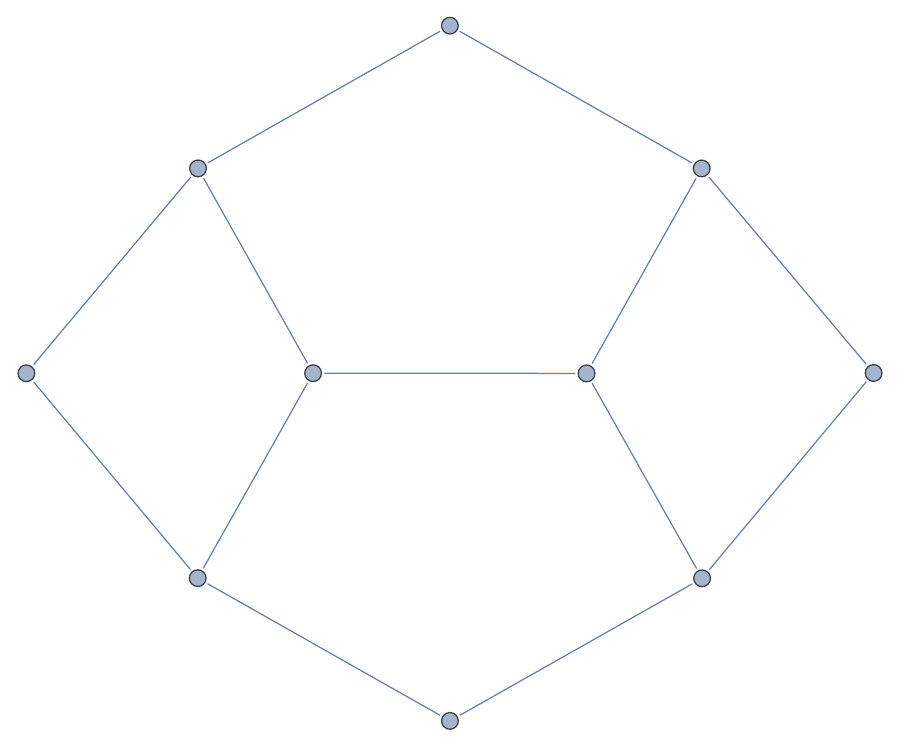} &\hspace{1cm}
  \includegraphics[width=6.5cm]{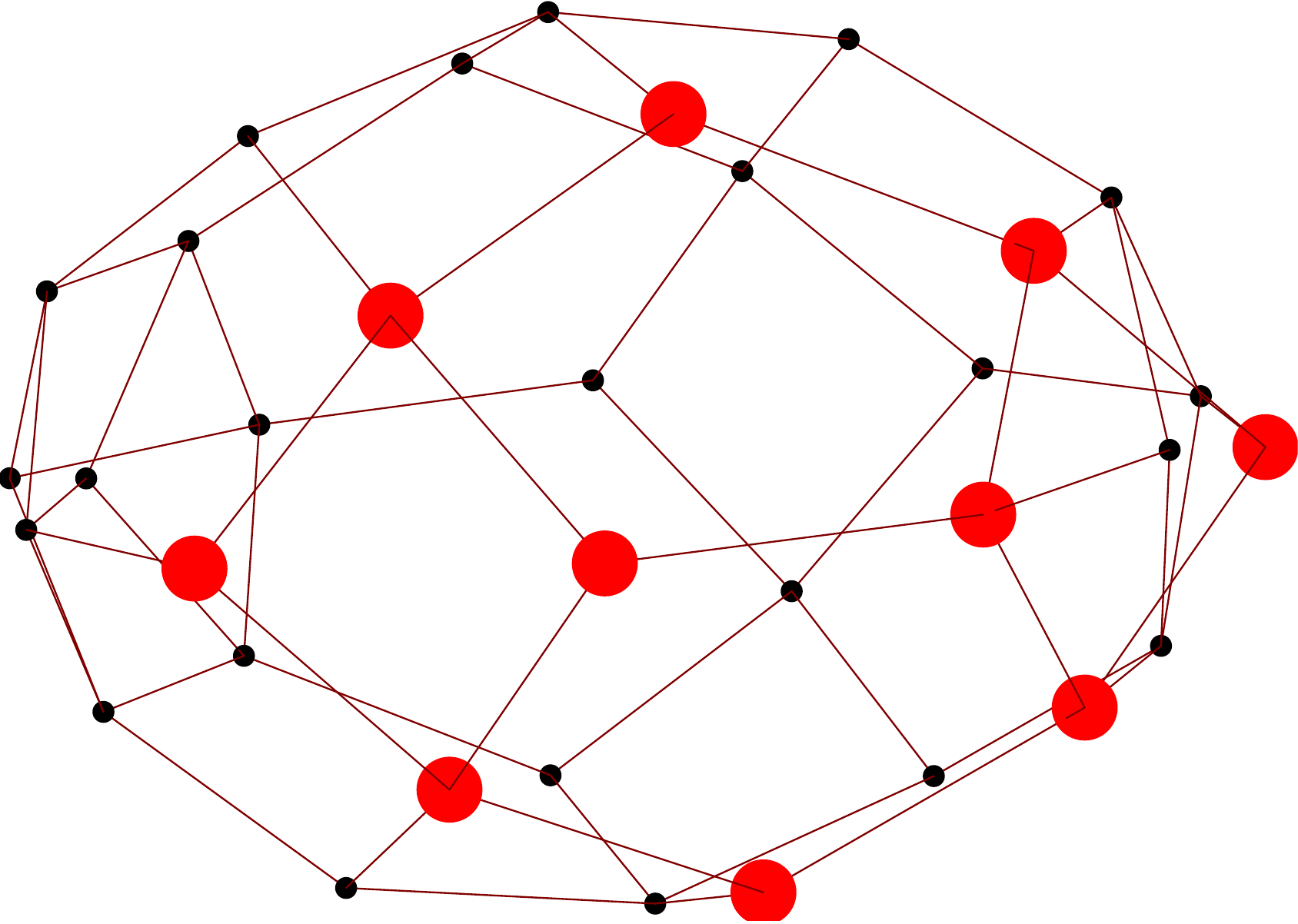}  
  \end{tabular}
  \caption{A plaquette and its embedding in the Faberg\'e egg.}
  \label{fig:plaquette}
\end{figure}

So much for the top-cell. Now we want to consider its first boundaries. There are six of these, and a particular one is represented by the plabic graph in figure \ref{fig:gr36bound}. It is the dual to the quiver which is obtained by freeze/deleting vertices $\langle 136\rangle/\langle 126\rangle$ from the Grassmannian quiver.
If we focus on the mutable nodes and their links, we see that mutating this seed leads to an $A_3$ cluster algebra -- as predicted in section \ref{ss:mhv}. However, unlike what happens in $Gr(2,6)$, the frozen vertices here look very different (they have to, since Pl\"ucker coordinates now have three entries, not two!), and this will imply that the algebra cannot be generated by mutating only nodes of valence four. Equivalently, the dual plabic graph has faces which are not squares. Therefore, the {\em accessible} algebra here is not a full $A_3$. Geometrically, the seeds and their links form a ``plaquette'' shown in figure \ref{fig:plaquette}. It is made up of two squares and two pentagons, and we should think of these as a subset of a full $A_3$ which has three squares and six pentagons. It is also a subset of the Faberg\'e egg, as shown in the same figure.

\begin{figure}[htpb]
  \centering
  \includegraphics[width=6.5cm]{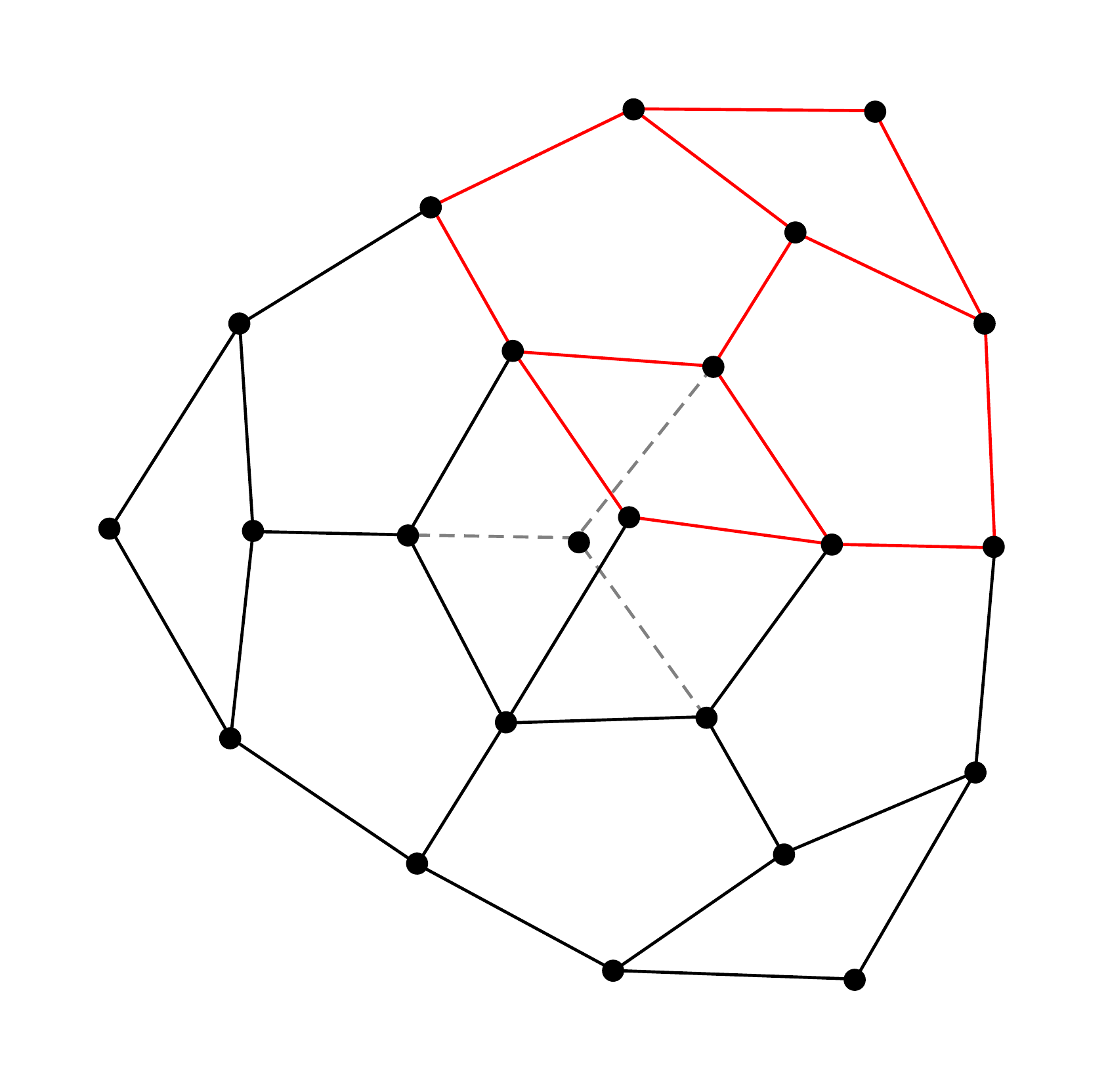}  
  \caption{One half of the Faberg\'e egg. It is almost covered completely by three plaquettes. One seed, corresponding to one of the corners of the cube, is lost when taking the first boundary. The story is the same for the other half.}
  \label{fig:halfegg}
\end{figure}
All six boundaries of the top cell graph of $Gr_{+}(3,6)$ can be found by taking the quiver \ref{fig:gr36topcell} and cyclically shifting the values in the angled brackets by one. In each of the resulting quivers, we freeze the same lower right vertex. Altogether, these six boundaries lead to six plaquettes which cover almost completely the Faberg\'e egg. Note: almost, but not quite. Two of the seeds are ``lost'' when taking boundaries. This is because those seeds are not attached to any of the six $A_3$ that are visible on the first boundary. This is illustrated in figure \ref{fig:halfegg}.

The NMHV 6 pt amplitude can be written in several ways, since there are several BCFW representations for it. The shortest representations have three terms only. If we label each of the six boundaries by the corresponding minor being set to zero, then there are two three-term representations corresponding to the boundaries $\langle 123\rangle, \langle 345\rangle, \langle 561\rangle$ and $\langle 234\rangle,\langle 456\rangle, \langle 612\rangle$. A natural question is to ask where the corresponding plaquettes lie. Remarkably, these two representations cover each one half of the Faberg\'e egg (except the two missing seeds mentioned before), as shown in figure \ref{fig:halfegg}. That is, these three term representations single out three particular $A_3$ subalgebras out of twelve possible ones, in such a way that each set covers a half of the accessible algebra. This is a refinement by a factor of two from the six subalgebras which are already singled out by the positroid stratification. Although we have been unable to determine it, it would be extremely interesting to determine whether the cluster algebra structure singles out these subsets in some way. In particular, cluster algebras of the form $Gr_{+}(3,n)$ with $n>8$ and $Gr_{+}(4,n)$ with $n>7$ have an infinite number of seeds, but clearly only a finite set containing only Pl\"ucker coordinates; out of these a further restricted set provides BCFW terms of scattering amplitudes. Which are these? Perhaps this can be understood from the intrinsic properties of the cluster algebra itself. It is likely that in order to understand this link we will require more data, in particular an analysis of higher point and higher $k$ tree amplitudes.

\bigskip
\acknowledgments{We would like to thank J.~Bourjailly, J.~Golden, M.~Rosso, M.~Spradlin, and A.~Volovich for discussions. Special thanks to S.~Brodsky for discussions and collaboration at the initial stages of this project.
M.P. is supported by DOE grant DE-SC0010010-Task A. BUWS is supported by the US Department of Energy under contract DE-FG02-11ER41742.}

\appendix


\section{Plabic Graphs and Cluster Algebras}
\label{sec:shell-graphs-cluster}
Here we review basic information on both plabic graphs and cluster algebras. We will be terse, and so recommend that the reader  interested in more details consult other references -- we have found particularly useful \cite{ArkaniHamed:2012nw}, \cite{Franco:2014nca} and \cite{2006math......9764P}

\subsection{Plabic graphs and the positive grassmannian}
\label{sec:shell-graphs}

A plabic graph is a \emph{pla}nar, \emph{bic}olored graph, built from three ingredients. Firstly, a set of vertices, which come in two colors -- black and white. Secondly, internal undirected edges which connect these vertices; and thirdly external edges which connect a vertex to a boundary; since graphs are planar, they are usually drawn on a disc. In particular this means they have unique dual graphs. In the following we will focus on so-called reduced plabic graphs, which for which the associated equivalence class contains no graph with double edges. We will also demand that all vertices have valency three. Each plabic graph so defined is representative of an associated decorated permutation $w:\mathbb{Z}\to\mathbb{Z}$, with 
\begin{align}
i\leq w(i) \leq i+n,\\
w(i+n) = w(i) + n
\end{align} 
To determine the permutation associated to a given graph we use the following construction  \cite{2006math......9764P}. We begin by labeling its external lines with numbers $1,\ldots,n$ in a counterclockwise fashion. Next we define left-right paths as follows: starting at a given exterior leg $i$, say, we move towards the interior of the graph. Everytime we reach a vertex, we turn: if it is white, we turn left, otherwise we turn right. In this way we will flow through the graph until we reach some other exterior leg $j$. If $j<i$ we set $j\to j+n$. In this way we have obtained the map $w$ above. The number
\bea
\frac{1}{n}\sum_{i=1}^{n}(w(i)-i) =k
\eea
counts the number of legs $i$ such that $w(i)>n$.

Now let us consider the equivalence class of reduced plabic graphs which lead to the same permutation. All such graphs can be obtained by starting from a single one and performing a set of moves. There are two kinds of moves: merge/unmerge and the so called square move. They have the property that graphs related by them lead to the same permutation.
\begin{figure}[htpb]
  \centering
  \begin{tabular}{c}
  \includegraphics[width=0.45\textwidth]{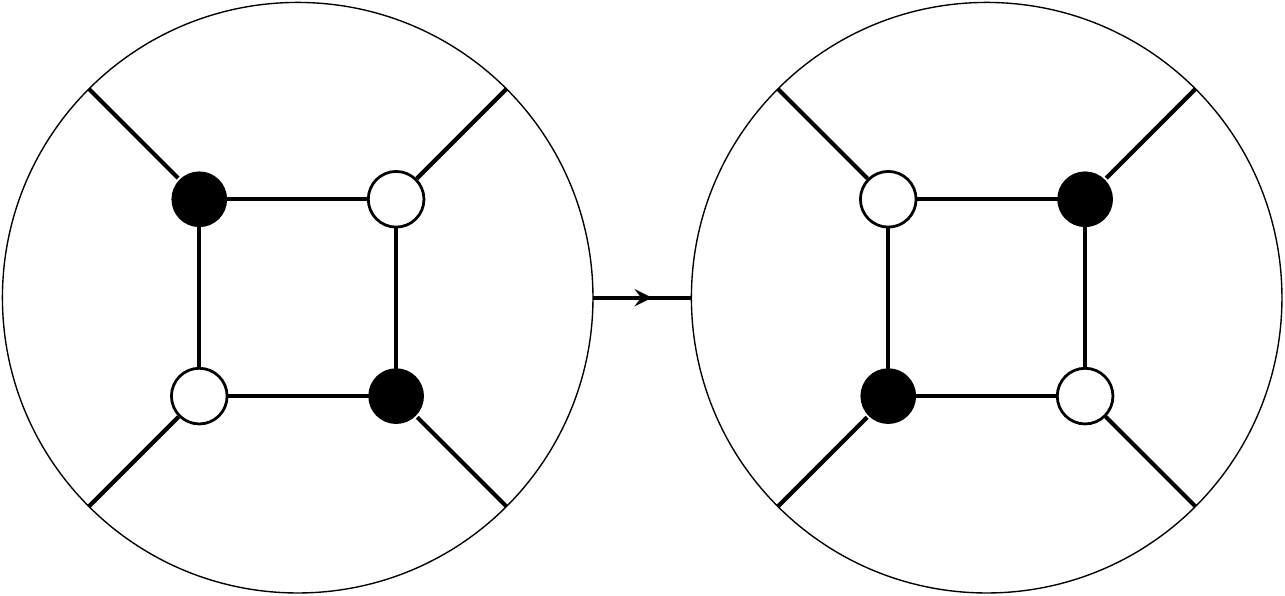}\\  \includegraphics[width=0.65\textwidth]{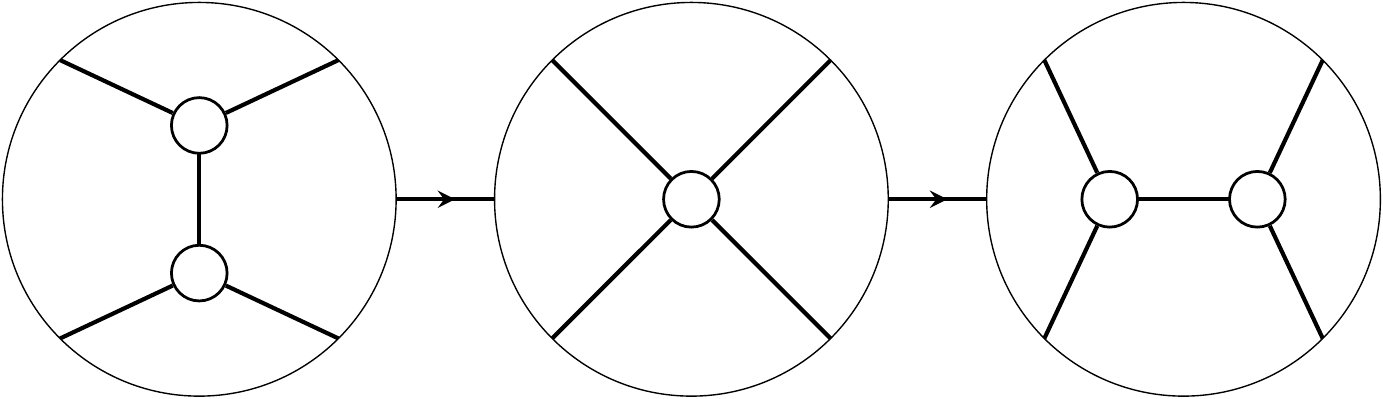}
  \end{tabular}

  \caption{The square move and mergers.}
  \label{fig:sqmvmerg}
\end{figure}
The square move is the most important, since it corresponds to a mutation in the cluster algebra associated to the graph. As shown in \cite{2006math......9764P} these equivalence classes of plabic graphs label cells in the positroid decomposition of the positive part of the Grassmannian $Gr_{+}(k,n)$, and it is to this object we turn next.

\subsection{The positroid decomposition}

The (real) Grassmannian $Gr(k,n)$ is the set of $k$-dimensional linear subspaces of $\mathds R^n$. It can be equivalently thought of as the coset $M(k,n)/GL(k)$. If we think of $k\times n$ matrices in terms of their $n$ columns $Z_i, i=1,\ldots n$, then a natural set of coordinates in the Grassmannian are ratios of $k\times k$ minors of the original matrix, i.e. objects such as
\bea
\frac{\langle i_1 \ldots i_k\rangle}{\langle j_1\ldots j_k\rangle}, \qquad \langle p_1 \ldots p_k\rangle \equiv \epsilon_{a_1\ldots a_k} Z^{a_1}_{p_1}\ldots Z^{a_k}_{p_k}.
\eea
The set of $\binom nk$ minors is not algebraically independent, since they satisfy Pl\"ucker relations. These in turn follow from contractions with the basic identities (Cramer's rule):
\bea
\sum_{\sigma} \mbox{sgn}(\sigma) \langle \sigma(i_1) \ldots \sigma(i_k)\rangle Z_{\sigma(i_{k+1})}^a=0,
\eea
for any set $\{i_1,\ldots i_{k+1}\}$. For instance, in $Gr(2,4)$ we have the single Pl\"ucker identity:
\bea
\langle 12\rangle \langle 34\rangle+\langle 23\rangle \langle 14\rangle=\langle 13\rangle \langle 24\rangle.
\eea
From this it follows that the Grassmannian can also be thought of as the projective algebraic variety defined in $\mathds P^{\binom nk}$ as the vanishing set of the Pl\"ucker relations. This is usually called the Pl\"ucker embedding of the Grassmannian in $\mathds P^{\binom nk}$, and the minors are usually called Pl\"ucker coordinates.

We are particularly interested in the positive part of the Grassmannian. This is defined as the subset obtained by demanding that all ordered Pl\"ucker coordinates should be non-negative, that is we demand
\bea
\langle i_1 \ldots i_k\rangle \geq 0 \quad \mbox{if} \quad i_1<i_2<\ldots <i_k
\eea
It follows from the Pl\"ucker relations that it is enough to specify which minors involving consecutive columns are positive \cite{2006math......9764P}. 
%
%
The positroid stratification decomposes the positive part of the Grassmannian into cells of different dimensionality. These cells are characterized by the linear relations among consecutive columns $Z_i$, or equivalently by which consecutive minors are vanishing, subject to the constraints imposed by the Pl\"ucker relations. Each such cell is an open algebraic subvariety of the original Grassmannian seen as a subset of $\mathds P^{\binom nk}$, and we will denote these as {\em positroid varieties}. For each cell we define, for each column $i$, $w(i)$ to be the first column $Z_{w(i)}$ such that $Z_{i}\in \mbox{span} \left\{ Z_{i+1},\ldots, Z_{w(i)}\right\}$. In this way to each positroid variety we have associated a permutation, and that in turn is associated to a plabic graph.

The stratification starts off with the {\em top-cell}, which is characterized by non-vanishing of all consecutive minors. The associated affine permutation is always given by $i\to i+k$. A plabic graph in the right equivalence class can be generated using Postnikov's method as explained in \cite{2006math......9764P}. We call this the \emph{top cell} diagram associated with $Gr_{+}(k,n)$. 
To obtain a higher codimension cell, we go to a boundary by sending some consecutive minor to zero. This always corresponds to deleting an edge in the associated graph, consequently reducing the number of faces. Since it is the number of faces that determines the dimensionality of the cell, this procedure systematically reduces it by one unit. At the same time, every time such an edge is removed it amounts to performing a transposition on the permutation labeling the graph, until we finally reach the identity permutation.

\subsection{Cluster algebras}
\label{sec:clusters}
Cluster algebras have only been discovered relatively recently \cite{MR1887642,MR2004457,MR2110627,MR2295199}. However, they have become a very actively researched field in recent years. For an introduction to different areas of research in cluster algebras see, {\em e.g.}, \cite{MR3119820} and \cite{website} . 

For our purposes, a cluster algebra is a commutative algebra $\mathcal A$ over some field $K$, equipped with a special collection of subsets denoted \emph{clusters}. The fundamental property of such clusters $(a_1,\ldots, a_{p})$ is that the full algebra $\mathcal A$ may be recovered by considering the set of Laurent polynomials in the cluster variables, $A\subset K[a_1^{\pm 1},\ldots,a_p^{\pm 1}]$. The cluster algebra structure of $\mathcal A$ is made up of two ingredients: a set of \emph{seeds} $\{Q,\{a_{i}\}\}$ -- each made of a quiver $Q$,  and a set of \emph{cluster $a$-coordinates} $\{a_{i}\}$ -- and a \emph{mutation rule} which relates different seeds. A quiver $Q$ can be represented by an adjacency matrix which we will denote by $q_{ij}$. Assuming that there are $p$ cluster ${\cal A}$-coordinates in a given seed, a cluster mutation relates a cluster coordinate in the seed $\{Q,\{a_{i}\}\}$ to a cluster coordinate in the seed $\{Q',\{a_{i}'\}\}$ via 
\bea 
a_{k}a_{k}' = \prod_{j=1}^{p}a_{j}^{\min(0,q_{kj})} + \prod_{j=1}^{p}a_{j}^{\min(0,-q_{kj})}.\label{eq:clusmut}
\eea
and
\bea
q'_{ij}=\left\{
\begin{tabular}{ll}
$-q_{ij}$, \quad & \mbox{if} $k\in \{i,j\} $\\
$q_{ij}$, \quad & \mbox{if} $q_{ik}q_{kj}\leq 0$\\
$q_{ij}\pm q_{ik}q_{kj}$, \quad & \mbox{if} $q_{ik},q_{kj}\gtrless 0$
\end{tabular}
\right. .
\eea
The mutation is more clearly expressed as the product over all cluster $a$-coordinates $a_{j}$ connected to $a_{i}$ with an arrow pointing from $j$ to $i$ plus the product of those with an arrow of the opposite orientation. The $a$ coordinates which are common to all seeds are said to be frozen -- by definition, one cannot mutate them to pass to a new seed. All other coordinates are said to be mutable.

The cluster structure of the algebra $\mathcal A$ can be determined systematically from an initial seed by mutating all possible unfrozen variables to determine a new batch of seeds, mutating those in turn, and so on. In general this procedure will never stop. However, there are special cluster algebras, the \emph{finite type} or \emph{ADE type} cluster algebras where this is the case. If at some point the mutable quiver (i.e. the part of the quiver pertaining to mutable variables) of some seed matches an ADE Dynkin diagram then the cluster algebra is finite. 
As shown by Scott in \cite{Scott2003}, the coordinate rings of the Grassmannians $Gr(2,n), Gr(3,6), Gr(3,7),Gr(3,8)$ are  examples of finite cluster algebras, with Pl\"ucker identities playing the role of mutation relations and cluster $a$-coordinates related to Pl\"ucker coordinates. In the ADE classification they correspond to $A_{n-3}, D_4,E_6$ and $E_8$ respectively.

We can attach two interesting geometrical objects to a cluster algebra. The first is the cluster simplicial complex. It is the polytope whose vertices are the cluster $a$-coordinates; its codimension-1 faces are the simplices corresponding to ensembles of coordinates belonging in a cluster. As an example, the cluster complex for the $D_4$ cluster algebra is shown in figure \ref{fig:duald4}. The second geometrical object is the dual of the cluster complex, which is usually called the generalized associahedron. It is the polytope obtained by taking for vertices the clusters (or seeds), and connecting two vertices by an edge if here is a mutation between the respective clusters. The generalized associahedron for the $A_3$ cluster algebra, with its 14 seeds, is shown in figure \ref{fig:a3poly}.

\bibliographystyle{utphys}
\bibliography{clusters,Biblio}

\end{document}